\documentclass[journal]{IEEEtran}
\usepackage{amsmath,amsfonts}
\usepackage{amssymb}
\usepackage{algorithmic}
\usepackage{array}
\usepackage[caption=false,font=footnotesize,labelfont=sf,textfont=sf,labelformat=simple]{subfig}
\usepackage{textcomp}
\usepackage{stfloats}
\usepackage{url}
\usepackage{verbatim}
\usepackage{graphicx}
\def\BibTeX{{\rm B\kern-.05em{\sc i\kern-.025em b}\kern-.08em
    T\kern-.1667em\lower.7ex\hbox{E}\kern-.125emX}}
\usepackage{balance}

\usepackage{booktabs}
\usepackage{array}
\usepackage{multirow}
\usepackage{cite}
\allowdisplaybreaks
\usepackage{xcolor}
\usepackage{placeins}

\begin{document}
\renewcommand{\thesubfigure}{(\alph{subfigure})}
\setlength{\abovedisplayskip}{4pt}
\setlength{\belowdisplayskip}{4pt}
\setlength{\abovedisplayshortskip}{2pt}
\setlength{\belowdisplayshortskip}{2pt}
\title{Structural Analysis and Internal Stability Enhancement of Virtual-Admittance-Based Cascaded GFMIs Under Unity Voltage-Feedback Decoupling}
\author{Stephen Arinze Obi,~%
        Ki-Hyun Kim,~\IEEEmembership{Graduate Student Member,~IEEE,}%
        ~and Jae-Jung Jung,~\IEEEmembership{Senior Member,~IEEE}%
\thanks{The authors are with the School of Electronics and Electrical Engineering,
Kyungpook National University, Daegu 41566, Republic of Korea
(e-mail: mashall96@knu.ac.kr; rlarlgus5615@knu.ac.kr; jj.jung@knu.ac.kr).}%
\thanks{Corresponding author: Jae-Jung Jung.}%
}

\markboth{}%
{}

\maketitle

\begin{abstract}
Virtual admittance (VA) is widely used in cascaded voltage-control and
current-control (VC-CC) grid-forming inverters (GFMIs) because it shapes
the converter terminal behavior while preserving the current-regulation
path required for current shaping and limiting. However, the achievable
VC-loop bandwidth remains strongly coupled to the CC-loop bandwidth and
to the VA parameters. Voltage-feedback decoupling (VFD) is commonly used
to relax this coupling, but in VA-based control its benefit is not
unconditional. This paper shows that unity-gain VFD, which represents the
full-decoupling condition, removes the low-frequency restoring term
associated with the filter capacitor and drives the voltage loop toward a
delay-sensitive double-integrator structure. This internal-stability
limitation is referred to here as the VFD trap. To address this trap
without attenuating VFD, a proportional active-damping (AD) path is
proposed, implemented as negative capacitor-voltage feedback in the
current-reference path. The proposed path restores the missing
low-frequency support while retaining unity VFD and introduces an
additional AD-based degree of freedom for VC-loop tuning. A minimum
support condition, a delay-aware phase-margin expression, and compact
forward/inverse design equations are derived for operating-point
selection. Standalone and grid-connected experiments on a 3-kVA prototype
verify the analysis, showing that the proposed path recovers stable
unity-VFD operation, reduces the voltage-step settling time from
approximately 9~ms to 3~ms, and maintains stable power injection.
\end{abstract}


\section{Introduction}

\IEEEPARstart{T}{he} growing share of inverter-based resources is
changing power-system dynamics and increasing the need for grid-forming
inverters (GFMIs) that provide voltage support, frequency support, and
inertia-like behavior without relying on a stiff
grid~\cite{Kroposki2017RenewableGrid, Lasseter2020GridFormingInverters}.
These functions are typically realized by an outer synchronization or
power-control layer (PCL) that establishes the converter voltage angle,
together with an inner voltage-control layer that regulates the output
voltage~\cite{Li2022RevisitingGridFormingFollowing,
Rosso2021GridFormingReview, Zhao2024LowFrequencyResonances}. Because
different inner-loop realizations lead to different dynamic limitations,
this work focuses on the cascaded voltage-control and current-control
(VC-CC) structure, which is widely used in practical inverter
implementations~\cite{Zhao2023InnerControlLoopsDamping,
Liu2024UnifiedVoltageControl, Liao2020PassivityVoltageControllers}. In
this structure, the inner voltage-control dynamics are a key design
constraint: insufficient bandwidth degrades voltage regulation, whereas
overly aggressive tuning can introduce high-frequency stability
problems~\cite{Li2022RevisitingGridFormingFollowing,
Wang2019HarmonicStability, Liao2019VoltageRegulatorsDualLoop,
He2025FeedforwardDampingPWMDelays}.
 
In cascaded VC-CC GFMIs, virtual admittance (VA) is commonly used to shape
the converter terminal behavior through a prescribed admittance. The VA
branch processes the voltage error between the internal voltage reference
and the measured capacitor voltage and generates the current reference
tracked by the inner current loop~\cite{Rodriguez2013VirtualAdmittance,
Wang2015VirtualImpedanceBasedControl,Wang2026GridFormingVirtualAdmittance,
Fan2022CurrentLimitingReview}. In this way, it preserves the current-regulation 
path required for current shaping and limiting. However, the achievable VC-loop 
dynamics remain strongly coupled to the CC-loop
bandwidth~\cite{Obi2025RevaluationVirtualAdmittance} and to the selected
VA parameters~\cite{Rodriguez2013VirtualAdmittance,
Wang2015VirtualImpedanceBasedControl, Huang2021ImpactVirtualAdmittance}.
This coupling is restrictive because the VA parameters are generally
selected for grid-support and power-loop
objectives~\cite{Huang2021ImpactVirtualAdmittance,
Cui2024VirtualAdmittanceClark}, whereas the CC-loop proportional gain is
limited by the filter inductance, LC-filter dynamics, and digital implementation
constraints~\cite{Akhavan2021PassivityEnhancement,
Holmes2009OptimizedCurrentRegulators}. Retuning the VA parameters to
accelerate the voltage loop is therefore not a free design choice. For
example, reducing the virtual inductance to raise the VC-loop crossover
moves it toward the LC-resonance region and reduces damping. As a result,
the VC-loop response cannot be assigned independently even when faster
voltage regulation is required.
 
Capacitor-voltage feedback decoupling, referred to here as voltage-feedback 
decoupling (VFD), is a common way to relax this bandwidth limitation in cascaded inverter control. 
By compensating the capacitor-voltage feedback path, VFD
extends the effective VC-loop bandwidth, reduces the direct dependence on
the CC loop, improves loop shaping, and mitigates LC-resonance-related
peaking while preserving the VC-CC
structure~\cite{Liao2019VoltageRegulatorsDualLoop,
He2025FeedforwardDampingPWMDelays, Zhao2023InnerControlLoopsDamping,
Liu2024UnifiedVoltageControl}. From an impedance-shaping viewpoint, it can
be interpreted as compensating the internal converter impedance so that
the inverter behaves closer to an ideal controlled voltage
source~\cite{Li2021ImpedanceCircuitModel}. However, in VA-based cascaded control, 
this benefit is not unconditional. Prior work has reported that a
near-unity VFD gain can destabilize the internal voltage loop, whereas
attenuating the gain below unity restores stable
operation~\cite{Obi2025RevaluationVirtualAdmittance}. The full-decoupling
condition that VFD is intended to provide therefore coincides with the
loss of internal stability, yet its structural cause has not been
explicitly established.
 
Existing remedies do not resolve this internal mechanism. The closest
prior work~\cite{Obi2025RevaluationVirtualAdmittance} recovers stability by
attenuating VFD, but this re-couples the
VC-loop dynamics to the current-loop bandwidth, reintroducing the
limitation that VFD was intended to remove. Related VA-based GFM-MMC
studies attribute instability to VA, current-loop, and feedforward
interactions and to negative-resistance regions under weak grids,
addressed through active damping~\cite{Gao2026VirtualAdmittanceGFMMMC},
while broader analyses of VC-loop feedforward in cascaded GFMIs adopt
feedforward attenuation as the remedial
action~\cite{Ravanji2023VoltageLoopFeedforward}. In each case, the remedy
either attenuates the decoupling or feedforward path itself or addresses
grid-side impedance interaction, rather than the internal, delay-driven
mechanism created by unity VFD. Consequently, the structural cause by which 
unity VFD removes the low-frequency capacitor support and reshapes the VC loop
has remained unexplained.
 
This paper addresses this gap. The unity-VFD limitation is shown to
originate from the removal of the low-frequency restoring term associated
with the filter capacitor, which drives the voltage loop toward a
delay-sensitive double-integrator structure; this internal limitation is
referred to as the VFD trap. To eliminate the trap without attenuating
VFD, a proportional active-damping (AD) path is proposed as negative
capacitor-voltage feedback in the current-reference path, restoring the
missing support while retaining unity VFD and providing an additional
AD-based degree of freedom for VC-loop tuning. The main contributions are
summarized as follows.
 
\begin{enumerate}
  \item A transfer-function framework is established to compare the
  baseline, attenuated-VFD, unity-VFD, and AD-supported unity-VFD
  configurations of VA-based GFMIs.
 
  \item Unity VFD is shown to remove the low-frequency restoring term
  associated with the filter capacitor, thereby producing the VFD trap as
  a delay-sensitive double-integrator behavior.
 
  \item A proportional AD path is proposed to restore the missing support
  while retaining unity VFD, thereby introducing an additional AD-based
  degree of freedom for VC-loop tuning.
 
  \item A minimum support condition, a delay-aware phase-margin
  expression, and compact forward/inverse design equations are derived and
  validated through full-model analysis and standalone/grid-connected
  experiments.
\end{enumerate}

\section{System Modeling and Structural Analysis}

This section derives the SISO small-signal model of the VA-CC GFMI to show
how VFD changes the VC-loop structure and then introduces the proposed
AD-based recovery. Since the mechanism of interest is internal to the
cascaded VC-CC layer, the slower PCL dynamics are neglected
\cite{Zhao2023InnerControlLoopsDamping}.

\subsection{Plant and Cascaded Control Structure}

\begin{figure}[!t]
    \centering
    \subfloat[]{%
        \includegraphics[width=0.95\columnwidth]{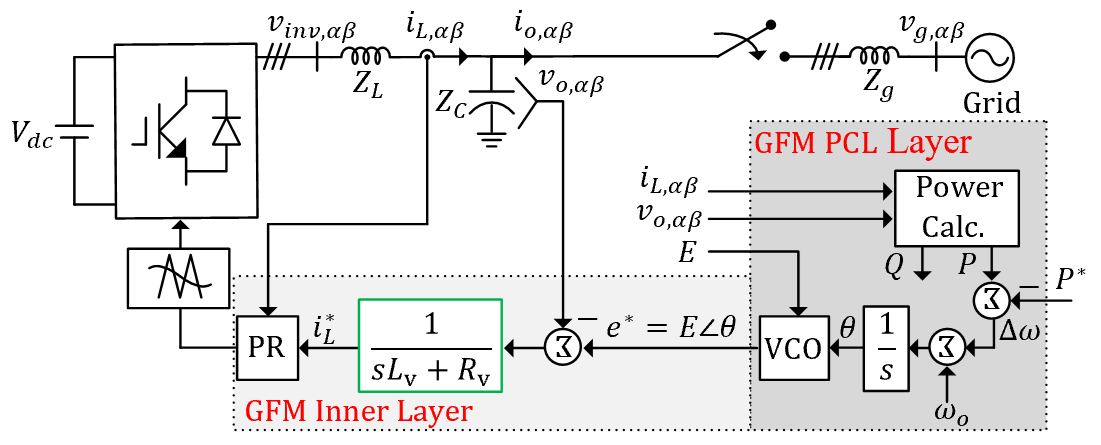}
        \label{fig:fig1a}}\\
    \subfloat[]{%
        \includegraphics[width=0.95\columnwidth]{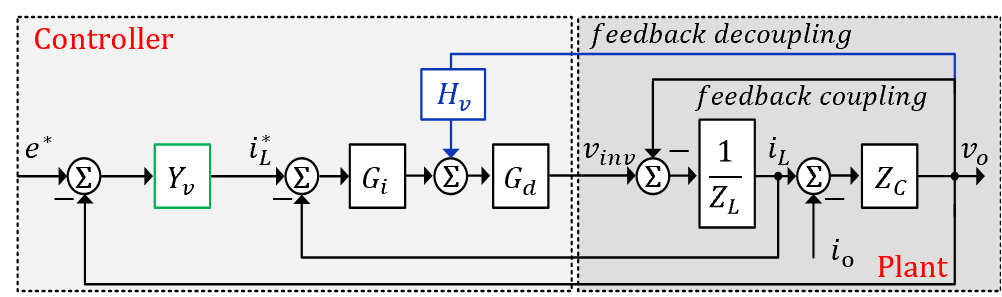}
        \label{fig:fig1b}}
    \caption{Single-line diagram of the VA-CC-VFD GFMI without the proposed
    AD path: (a) overall system structure and (b) simplified equivalent
    inner-loop control structure.}
    \label{fig:fig1}
    \vspace{-15pt}
\end{figure}

\begin{table}[!t]
\caption{Common System and Control Parameters}
\label{tab:parameters}
\centering
\renewcommand{\arraystretch}{1.15}
\begin{tabular}{c c c c}
\toprule
\toprule
Symbol & Description & Value (SI) & p.u. \\
\midrule
$S_b$ & Rated power & 3 kVA & 1.0 \\
$V_n$ & Rated voltage (rms) & 110 V & 1.0 \\
$f_o$ & Fundamental frequency & 50 Hz & 1.0 \\
$Z_b$ & Base impedance & $4.033~\Omega$ & 1.0 \\
$L_f,C_f$ & LC filter & 1 mH, $30~\mu$F & -- \\
$f_{sw},f_s$ & Switching/Sampling freq. & 10 kHz, 10 kHz & -- \\
$L_v,R_v$ & VA parameters & 4.49 mH, $0.47~\Omega$ & -- \\
$K_{psc}$ & PSC gain & $2.1\times 10^{-3}$ & -- \\
$H_v$ & VFD gain & 1 & -- \\
$K_r$ & CC resonant gain & $80.66~\Omega$ & 20 \\
$K_p,K_{ad}$ & Tuning variables & Variable $\Omega$, S & -- \\
\bottomrule
\bottomrule
\end{tabular}
\vspace{-10pt}
\end{table}

Fig. \ref{fig:fig1} shows the control structure of the considered
three-phase LC-filtered VA-CC GFMI for standalone and grid-connected
operation. The controller hierarchy consists of an outer
PCL and an inner cascaded VC-CC layer. Within the cascaded layer, the VC
loop is the outer voltage loop, and the CC loop is the inner current loop. The
PCL, implemented with an active-power synchronization controller (PSC),
provides the inverter angle $\theta$ for frequency-forming operation, while
the internal voltage magnitude reference $E$ is set by the nominal voltage
$V_n$ for voltage-forming operation \cite{Zhang2010PowerSynchronization}.
Here, $P$ and $Q$ denote the active and reactive powers, respectively, and
$P^\ast$ denotes the active-power reference. The resulting voltage reference
is expressed as $e^\ast=E\angle\theta$.

The inner control layer employs a cascaded VC-CC structure consisting of a
VA-based outer VC loop, an inner CC loop, and a capacitor-voltage feedback
decoupling path as in Fig. \ref{fig:fig1b}. These inner-loop controllers are
implemented in the stationary $\alpha\beta$-frame. For a balanced three-phase
system, the $\alpha$- and $\beta$-axes can be treated as decoupled
single-phase systems
\cite{Liao2019VoltageRegulatorsDualLoop,He2025FeedforwardDampingPWMDelays}.
Therefore, the following derivation is presented for one equivalent axis
under the balanced-system assumption. The physical plant is formed by the
filter inductance $L_f$ and the filter capacitance $C_f$. To isolate the
structural stability mechanism, parasitic resistances are neglected.

For one equivalent axis, the unloaded LC plant is represented by
$Z_L(s)=sL_f$ and $Z_C(s)=1/(sC_f)$, with $v_o=Z_C(s)i_L$. The unloaded
condition is used to expose the intrinsic internal-loop mechanism of the
VA-based inner-loop structure. The VA-based outer voltage controller is

\begin{equation}
Y_v(s)=\frac{1}{sL_v+R_v},
\label{eq:Yv}
\end{equation}
where $L_v$ and $R_v$ are the virtual inductance and resistance,
respectively. The current reference is given by
$i_L^\ast=Y_v(s)(e^\ast-v_o)$, where $e^\ast$ denotes the internal voltage
reference generated by the PCL. The inner current controller is implemented
in PR form as

\begin{equation}
G_i(s)=K_p+\frac{2K_r\omega_i s}{s^2+2\omega_i s+\omega_o^2},
\label{eq:Gi}
\end{equation}
where the proportional gain $K_p$ sets the nominal CC-loop crossover as
$K_p=\omega_{c,i}L_f$ \cite{Holmes2009OptimizedCurrentRegulators}, $K_r$ is
the resonant gain, $\omega_i$ is the resonant bandwidth
($\omega_i=\pi$ selected to accommodate a 1-Hz frequency-drift range), and
$\omega_o=2\pi f_o$ is the fundamental angular frequency.

The digitally controlled inverter voltage output is

\begin{equation}
v_{inv}
=
G_d(s)\left[G_i(s)(i_L^\ast-i_L)+H_vv_o\right],
\label{eq:vinv}
\end{equation}
where $H_v$ is the VFD gain and $G_d(s)=e^{-sT_d}$ models the digital control
delay. For a system sampled at $f_s$, with $T_s=1/f_s$, the total delay is
generally $T_d=1.5T_s$ \cite{Liao2019VoltageRegulatorsDualLoop}. The common
system and control parameters used for analysis and experiments are
summarized in Table \ref{tab:parameters}.

Substituting the plant and control relations gives the full-model open-loop
VC transfer function

\begin{equation}
\begin{aligned}
T_v(s)
&=
\frac{G_d(s)G_i(s)Y_v(s)}
{s^2L_fC_f+sC_fG_d(s)G_i(s)+1-G_d(s)H_v}  \\
&=
\frac{G_d(s)G_i(s)Y_v(s)}{\Delta_i(s)} .
\end{aligned}
\label{eq:Tv}
\end{equation}
The corresponding closed-loop VC transfer function is

\begin{equation}
T_{v,cl}(s)=\frac{T_v(s)}{1+T_v(s)} .
\label{eq:Tvcl}
\end{equation}

Hereafter, the full model denotes the unreduced SISO small-signal
transfer-function model for the configuration under consideration. For the
baseline and VFD cases, this corresponds to \eqref{eq:Yv}--\eqref{eq:Tvcl};
after AD is introduced, it refers to the corresponding AD-augmented model
before low-frequency reduction. The structural analysis uses leading-order
approximations around the VC-loop crossover: $Y_v(s)\approx1/(sL_v)$,
$G_i(s)\approx K_p$ with $K_p\gg sL_f$, and $G_d(s)\approx1$, where
applicable. The resonant branch of $G_i(s)$ is omitted only in the structural
derivation because the mechanism of interest lies well above $f_o$; it is
retained in the full-model analysis.

Here, the VC-loop bandwidth refers to the open-loop VC-loop crossover
frequency, with $f_{c,v}=\omega_{c,v}/(2\pi)$ and
$f_{c,i}=\omega_{c,i}/(2\pi)$ denoting the VC- and CC-loop crossover
frequencies, respectively. The transfer functions $T_v^{\mathrm{B}}(s)$,
$T_v^{\mathrm{T}}(s)$, $T_v^{\mathrm{A}}(s)$, and
$T_v^{\mathrm{AD}}(s)$ denote the baseline, unity-VFD trap, attenuated-VFD,
and unity-VFD-with-AD cases, respectively.

\subsection{Structural Effect of VFD}

This subsection compares the baseline, unity-VFD, and attenuated-VFD
structures to show how VFD modifies the low-frequency restoring term.

When VFD is disabled, $H_v=0$, and the VC transfer function becomes

\begin{equation}
T_v^{\mathrm{B}}(s)
=
\frac{G_d(s)K_p}
{sL_v\left[s^2L_fC_f+sC_fK_pG_d(s)+1\right]} .
\label{eq:Tv_baseline}
\end{equation}
The constant term in the denominator represents the low-frequency restoring
contribution of the filter capacitor. Around the VC-loop crossover where
$G_d(s)\approx1$, the leading-order bandwidth relation is

\begin{equation}
\omega_{c,v}^{\mathrm{B}}\approx\frac{K_p}{L_v}
=
\omega_{c,i}\frac{L_f}{L_v},
\label{eq:wc_baseline}
\end{equation}
and the corresponding phase margin is approximately

\begin{equation}
\mathrm{PM}\approx90^\circ-\omega_{c,v}^{\mathrm{B}}T_d\frac{180^\circ}{\pi}.
\label{eq:pm_baseline}
\end{equation}
Thus, the baseline VA-CC GFMI is structurally stable, but its VC-loop
bandwidth remains tied to the CC-loop bandwidth and the VA parameters.

To relax this bandwidth coupling and enhance dynamic performance, VFD is
introduced through the capacitor-voltage feedback path. For unity VFD,
$H_v=1$. Substituting $H_v=1$ and $G_d(s)\approx1$ into \eqref{eq:Tv},
the term $1-G_d(s)H_v$ in $\Delta_i(s)$ evaluates to $1-1=0$, removing
the low-frequency restoring constant from the denominator. In the intended cascaded
range, $\omega_{c,v}<\omega_{c,i}$, the current loop is effectively faster
than the voltage loop around the VC-loop crossover, and $sL_f\ll K_p$.
Hence, the open-loop VC transfer function reduces to

\begin{equation}
T_v^{\mathrm{T}}(s)\approx\frac{1}{s^2L_vC_f}.
\label{eq:Tv_trap}
\end{equation}
The corresponding trap-frequency estimate is

\begin{equation}
\omega_{trap}\approx\frac{1}{\sqrt{L_vC_f}}.
\label{eq:wtrap}
\end{equation}

From \eqref{eq:Tv_trap}, unity VFD removes the low-frequency restoring term
associated with the filter capacitor and drives the VC loop toward a
double-integrator form. Once control delay is included, the loop already
contributes $-180^\circ$, while the delay adds $-\omega_{trap}T_d$ (i.e.,
$f_{trap}=\omega_{trap}/(2\pi)$). Hence, the approximate phase margin near
the trap frequency is

\begin{equation}
\mathrm{PM}\approx-\omega_{trap}T_d\frac{180^\circ}{\pi}.
\label{eq:pm_trap}
\end{equation}
Thus, unity VFD relaxes the direct VC-CC bandwidth coupling only by
eliminating the restoring term that stabilizes the original plant, making
the loop highly delay-sensitive. This delay-sensitive double-integrator
behavior is referred to here as the VFD trap.

A practical remedy is to attenuate the VFD gain by selecting
$0<H_v<1$ \cite{Obi2025RevaluationVirtualAdmittance}. In that case, the
denominator of the inner loop retains the nonzero term $1-H_v$, giving

\begin{equation}
T_v^{\mathrm{A}}(s)\approx\frac{K_p}{sL_v(1-H_v)}.
\label{eq:Tv_attenuated}
\end{equation}
The resulting VC-loop crossover is

\begin{equation}
\omega_{c,v}^{\mathrm{A}}
\approx
\frac{K_p}{L_v(1-H_v)}
=
\frac{\omega_{c,i}L_f}{L_v(1-H_v)},
\label{eq:wc_attenuated}
\end{equation}
and the corresponding phase margin is

\begin{equation}
\mathrm{PM}\approx90^\circ-\omega_{c,v}^{\mathrm{A}}T_d\frac{180^\circ}{\pi}.
\label{eq:pm_attenuated}
\end{equation}

This explains why attenuated VFD can recover stability. However,
\eqref{eq:wc_attenuated} shows that this recovery is obtained by weakening
the VFD action and re-coupling the VC-loop crossover to $K_p$, and therefore
to the CC-loop bandwidth. Hence, the main benefit of VFD is only partially
retained.

\subsection{Proposed Structural Recovery via AD}

Section II-B showed that unity VFD removes the low-frequency support needed
to stabilize the VA loop. A proportional AD path is therefore proposed,
implemented as negative capacitor-voltage feedback in the current-reference
path:

\begin{equation}
i_L^\ast=Y_v(s)(e^\ast-v_o)-K_{ad}v_o .
\label{eq:ilref_ad}
\end{equation}

\begin{figure}[!t]
    \centering
    \includegraphics[width=0.95\columnwidth]{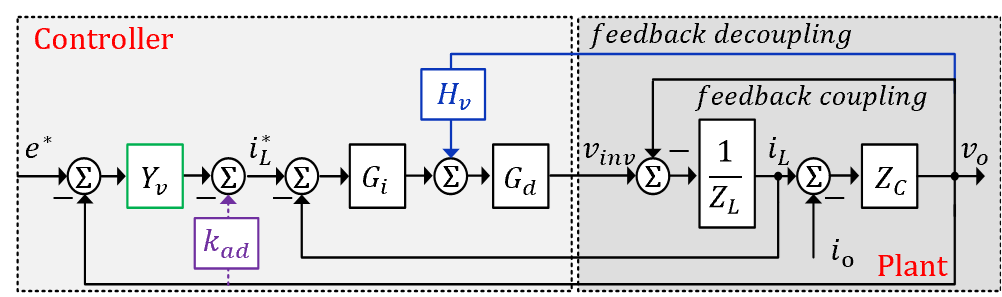}
    \caption{Simplified equivalent inner-loop control structure for the
    VA-CC-VFD GFMI with the proposed AD path.}
    \label{fig:fig2}
    \vspace{-15pt}
\end{figure}

\begin{table}[!t]
\caption{Structural Comparison of the Considered VA-CC GFMI Configurations:
Low-Frequency Approximations}
\label{tab:structural_comparison}
\centering
\renewcommand{\arraystretch}{1.15}
\begin{tabular}{c c c}
\toprule
\toprule
Config. & Support term & Stability/VC tuning \\
\midrule
Baseline VA-CC & $1$ & Stable; CC-coupled \\
Unity-VFD trap & $0$ & Decoupled; unstable \\
Attenuated-VFD & $1-H_v$ & Stable; recoupled \\
Unity VFD with AD & $K_pK_{ad}$ & Stable; $K_{ad}$-tuned \\
\bottomrule
\bottomrule
\end{tabular}
\vspace{-10pt}
\end{table}

The corresponding AD-augmented equivalent inner-loop structure is shown in
Fig. \ref{fig:fig2}. With AD, the inner-loop denominator
$\Delta_{ad}(s)$ adds the low-frequency support term $K_pK_{ad}$ to
$\Delta_i(s)$ in \eqref{eq:Tv}. Therefore, under unity VFD, the open-loop VC
transfer function reduces to

\begin{equation}
T_v^{\mathrm{AD}}(s)
\approx
\frac{K_p}{sL_v(K_pK_{ad})}
=
\frac{1}{sL_vK_{ad}}.
\label{eq:Tv_ad}
\end{equation}
The corresponding VC-loop crossover frequency becomes

\begin{equation}
\omega_{c,v}^{\mathrm{AD}}\approx\frac{1}{L_vK_{ad}},
\label{eq:wc_ad}
\end{equation}
and the associated phase margin is

\begin{equation}
\mathrm{PM}\approx90^\circ-\omega_{c,v}^{\mathrm{AD}}T_d\frac{180^\circ}{\pi}.
\label{eq:pm_ad}
\end{equation}

Equations \eqref{eq:wc_attenuated} and \eqref{eq:wc_ad} show that
attenuated VFD and unity VFD with AD restore the missing low-frequency
support through equivalent reduced terms when $K_pK_{ad}=1-H_v$. The special
case $K_pK_{ad}=1$ therefore corresponds to the baseline-support level and
gives $\omega_{c,v}^{\mathrm{AD}}\approx K_p/L_v$. This equivalence is only
structural: attenuated VFD restores support by reducing $H_v$, whereas the
proposed method keeps $H_v=1$ and restores support through a separate AD
path. Thus, $K_{ad}$ provides the primary VC-loop tuning variable under unity
VFD, although delay, retained filter dynamics, and finite current-loop
bandwidth introduce residual $K_p$-dependence.

Although \eqref{eq:wc_ad} shows how $K_{ad}$ shapes the VC-loop bandwidth,
$K_{ad}$ cannot be chosen arbitrarily small. Applying a first-order delay
approximation to the VC characteristic polynomial and setting its leading
low-frequency coefficient to zero (see Appendix) yields the minimum
support condition:

\begin{equation}
K_{ad,\min}=\frac{T_d}{L_v}.
\label{eq:kad_min}
\end{equation}

Equation \eqref{eq:kad_min} gives the minimum low-frequency support required
to avoid the unity-VFD trap. With the parameters in Table \ref{tab:parameters},
it yields $K_{ad,\min}=0.0334~\mathrm{S}$, or
$K_{ad,\min,\mathrm{pu}}\approx0.14$. However, this support bound alone does
not guarantee sufficient dynamic robustness or a desired VC-loop bandwidth;
it defines the minimum admissible $K_{ad}$.

The structural implications of the four main VA-CC GFMI configurations are
summarized in Table \ref{tab:structural_comparison}. The table highlights
that attenuated VFD and unity VFD with AD both restore low-frequency support,
but only the AD-based case retains unity VFD while making $K_{ad}$ the
primary VC-loop tuning variable.

\section{Analytical Validation and Design Interpretation}

This section validates the structural conclusions of Section II using the
corresponding full model, identifies the admissible tuning region, and derives
compact equations for operating-point selection.

\subsection{Structural Bode Validation and Design Region}

\begin{figure}[!t]
    \centering
    \includegraphics[width=0.95\columnwidth]{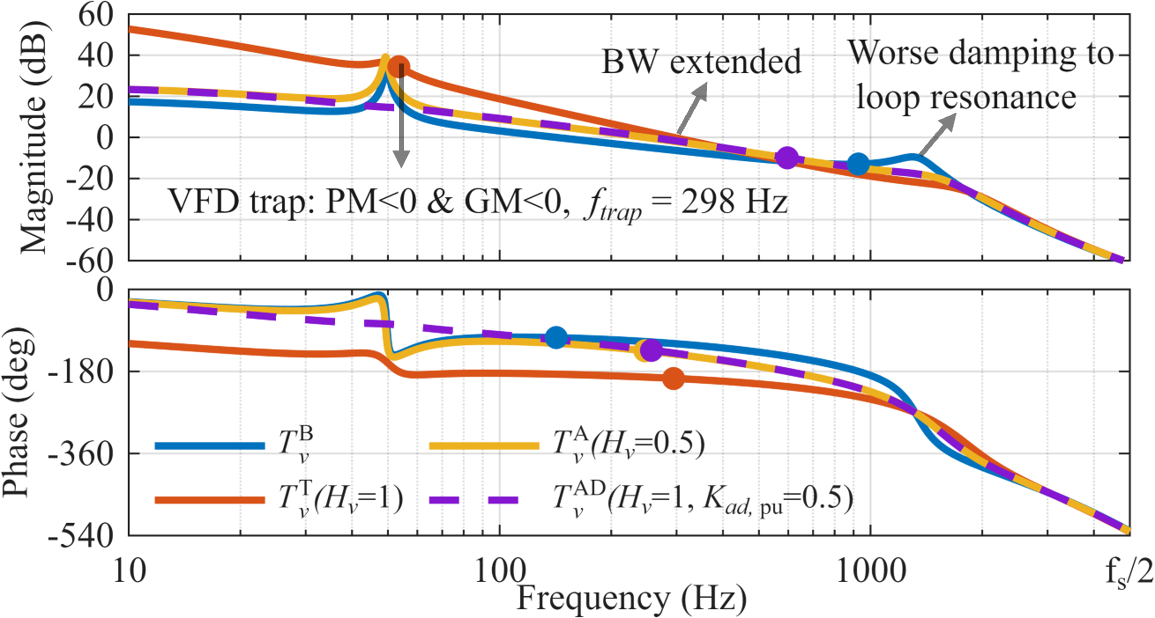}
    \caption{Open-loop Bode comparison of the baseline VA-CC GFMI,
    unity-VFD trap case, attenuated-VFD workaround, and unity VFD with AD at
    $K_{p,\mathrm{pu}}=1$.}
    \label{fig:fig3}
    \vspace{-15pt}
\end{figure}

Fig. \ref{fig:fig3} compares the four configurations using full-model Bode
responses for $K_{p,\mathrm{pu}}=1.0$, corresponding to
$f_{c,i}\approx642~\mathrm{Hz}$. The baseline case remains stable but
bandwidth-limited, while VFD extends the VC-loop bandwidth and mitigates the
resonance-related peaking observed in the baseline response. However, unity
VFD without AD exhibits the VFD-trap behavior consistent with the mechanism
in \eqref{eq:Tv_trap}--\eqref{eq:pm_trap}, losing stability near the
full-model trap region. Both attenuated VFD and unity VFD with AD restore
support, but only the AD-based case preserves unity VFD while using $K_{ad}$
for VC-loop shaping.

\begin{figure}[!t]
    \centering
    \subfloat[]{%
        \includegraphics[width=0.48\columnwidth]{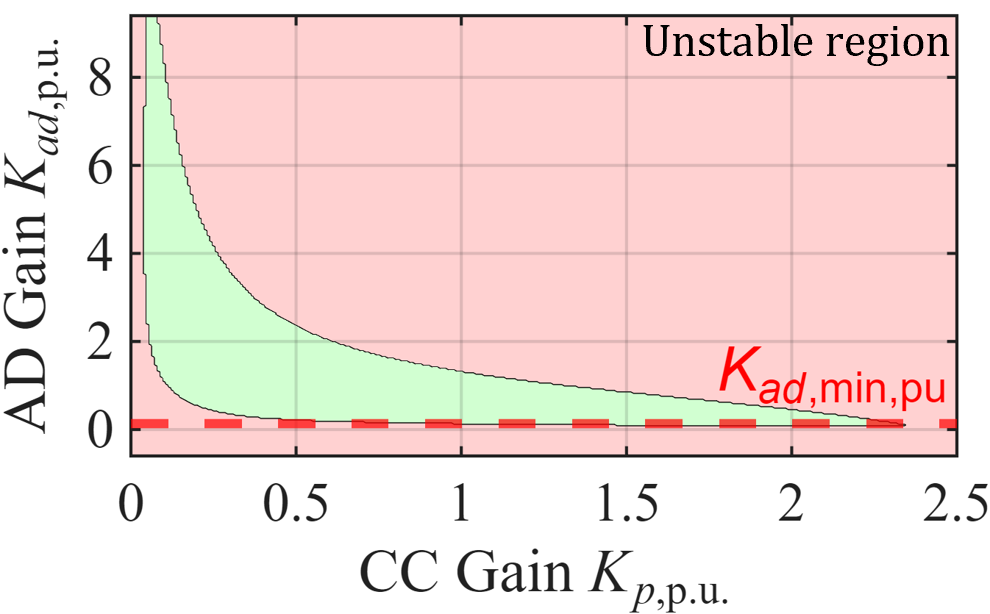}
        \label{fig:fig4a}}
    \hfill
    \subfloat[]{%
        \includegraphics[width=0.48\columnwidth]{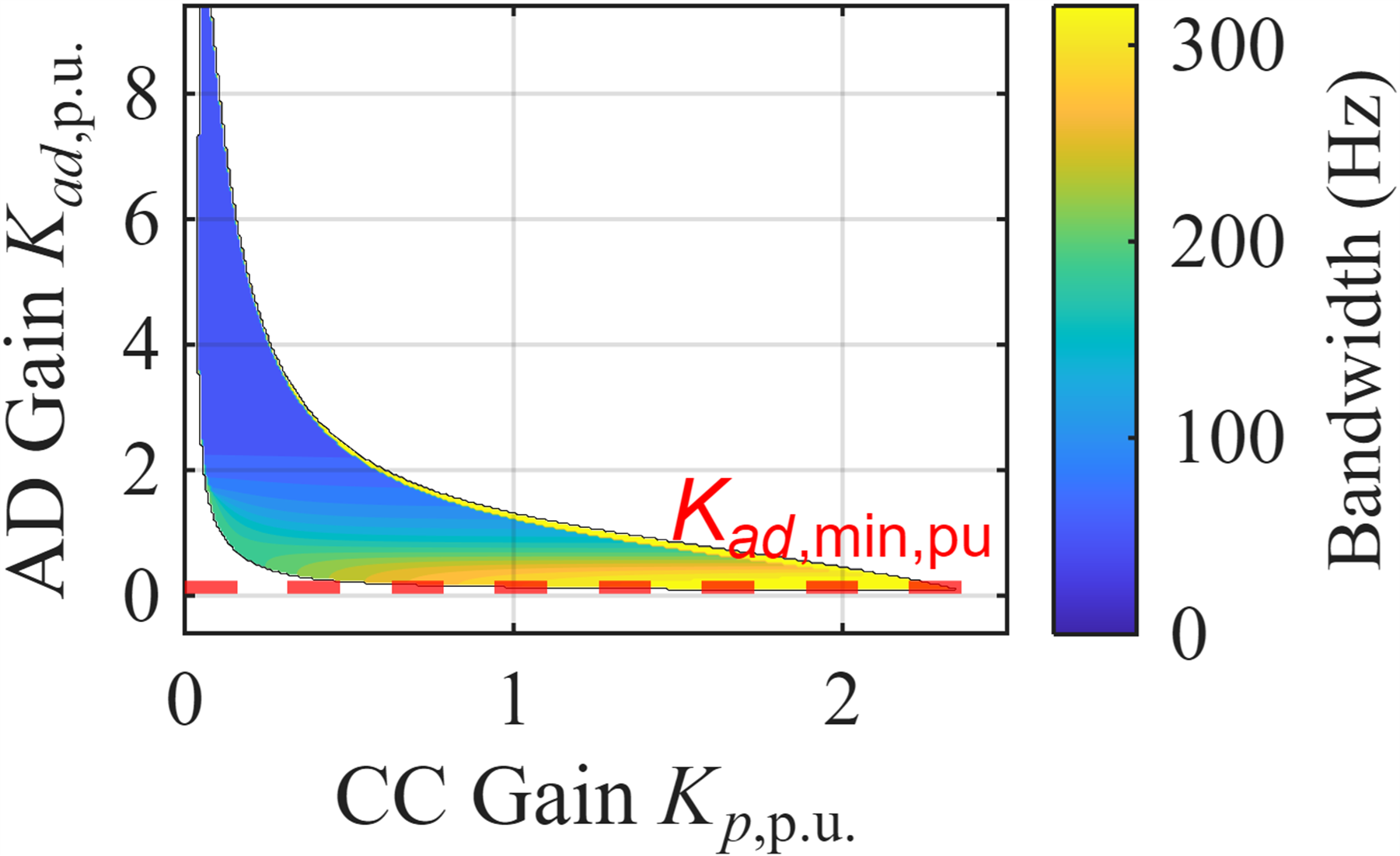}
        \label{fig:fig4b}}\\
    \subfloat[]{%
        \includegraphics[width=0.48\columnwidth]{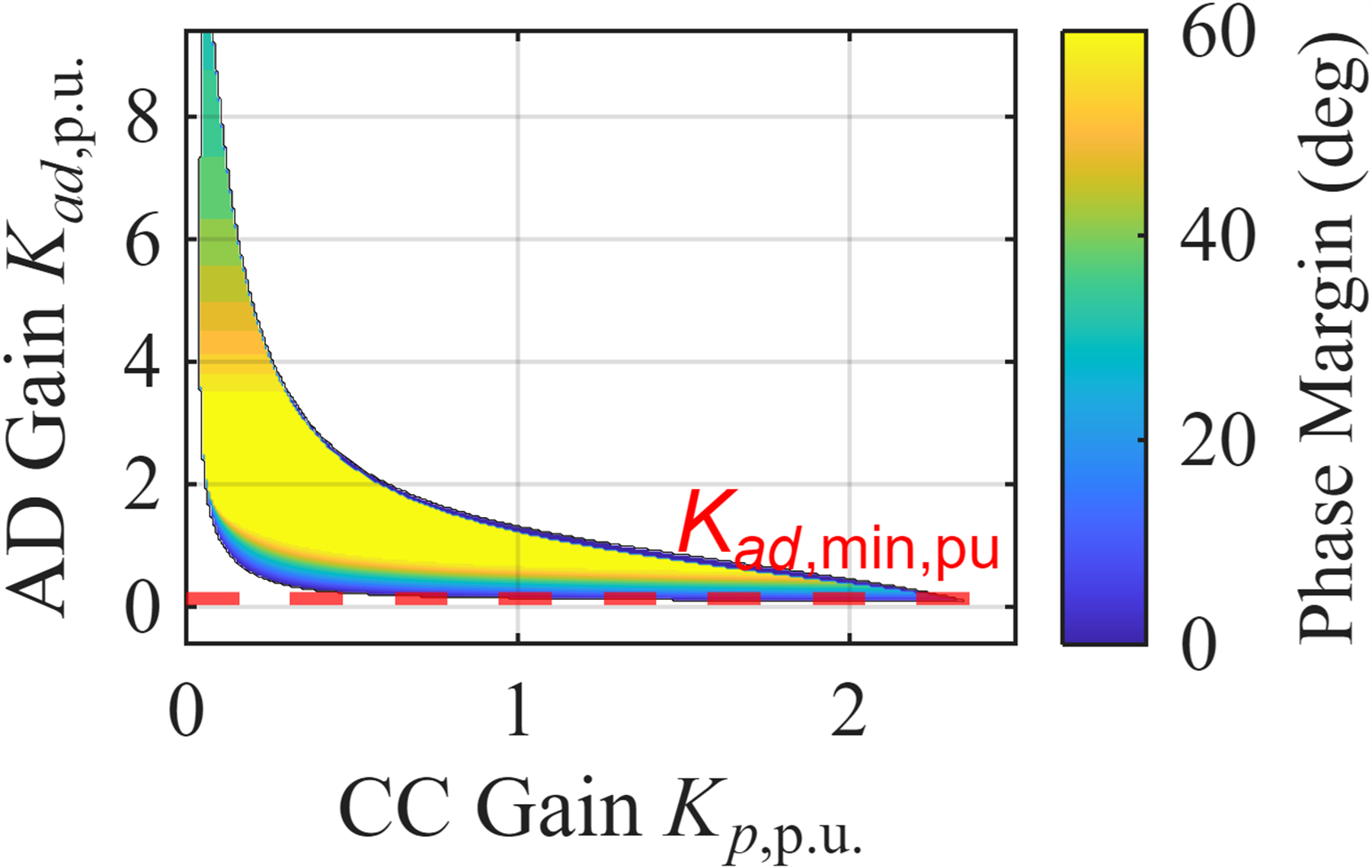}
        \label{fig:fig4c}}
    \hfill
    \subfloat[]{%
        \includegraphics[width=0.48\columnwidth]{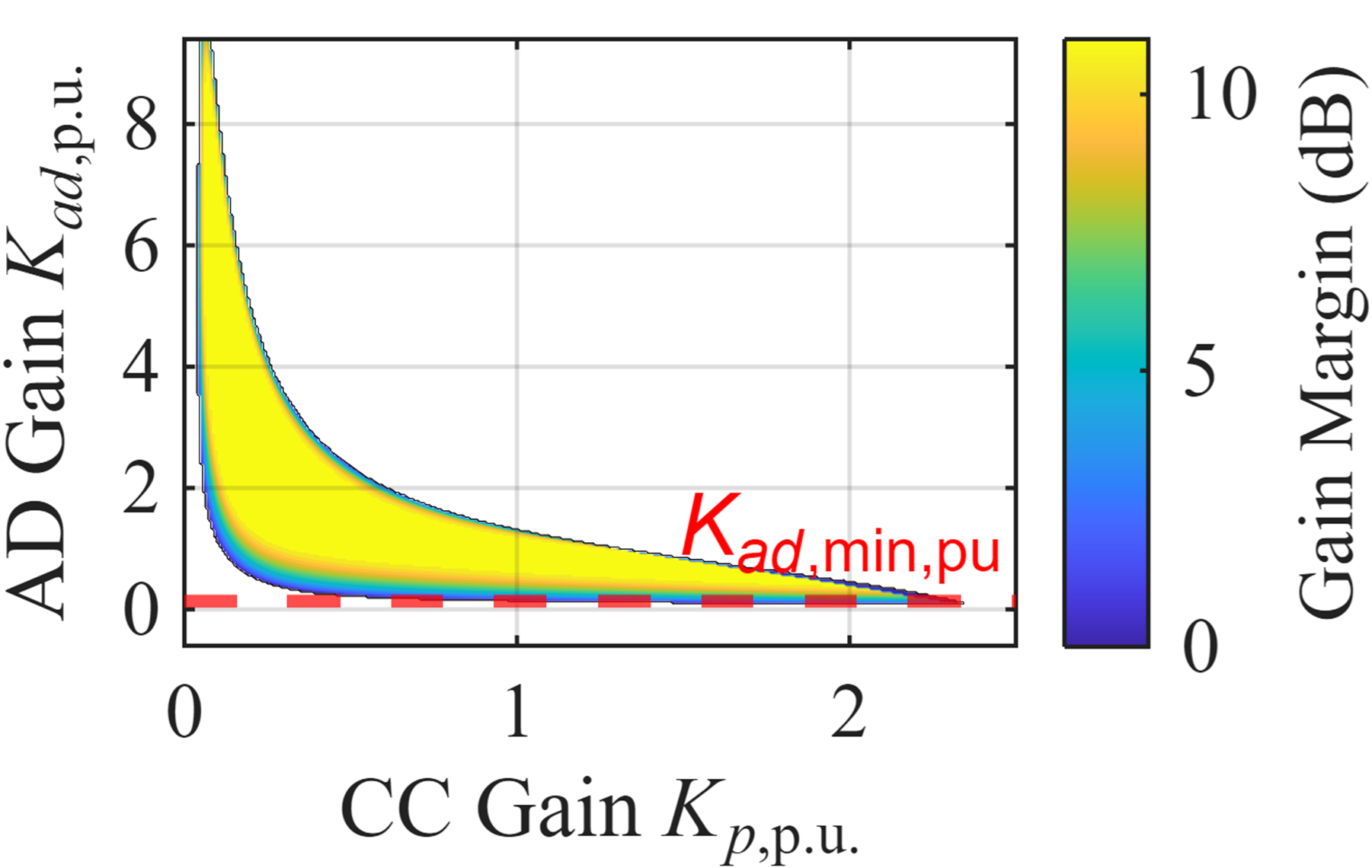}
        \label{fig:fig4d}}
    \caption{Full-model design maps in the $(K_{p,\mathrm{pu}},
    K_{ad,\mathrm{pu}})$ plane: (a) stable tuning region, (b) VC-loop
    bandwidth, (c) gain margin, and (d) phase margin, with the support floor
    indicated at $K_{ad,\min,\mathrm{pu}}\approx0.14$.}
    \label{fig:fig4}
    \vspace{-15pt}
\end{figure}

Fig. \ref{fig:fig4} summarizes the full-model design space of the proposed
unity VFD with AD in the $(K_{p,\mathrm{pu}},K_{ad,\mathrm{pu}})$ plane.
Fig. \ref{fig:fig4a} shows the stable tuning region and the analytical
support floor $K_{ad,\min}=T_d/L_v$, while Figs. \ref{fig:fig4b}--(d) show
the corresponding VC-loop bandwidth, gain margin, and phase margin. These
maps identify the useful tuning range where stability, bandwidth, and
robustness are jointly acceptable.

\begin{figure}[!t]
    \centering
    \subfloat[]{%
        \includegraphics[width=0.48\columnwidth]{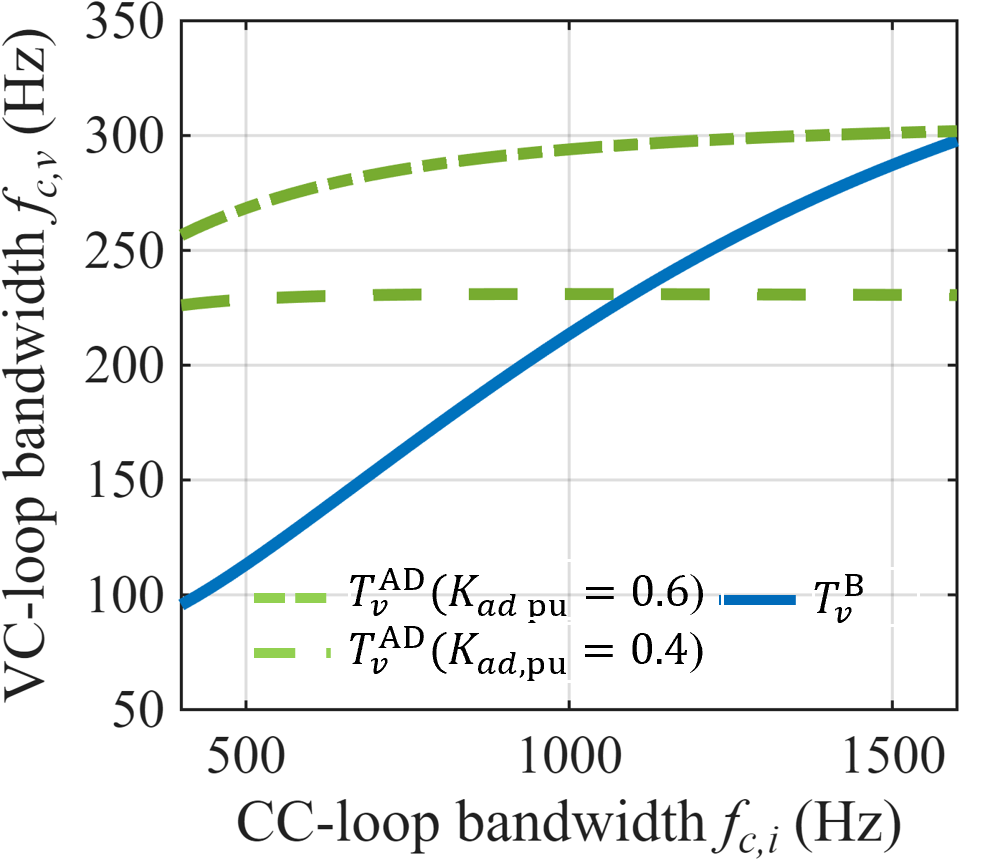}
        \label{fig:fig5a}}
    \hfill
    \subfloat[]{%
        \includegraphics[width=0.48\columnwidth]{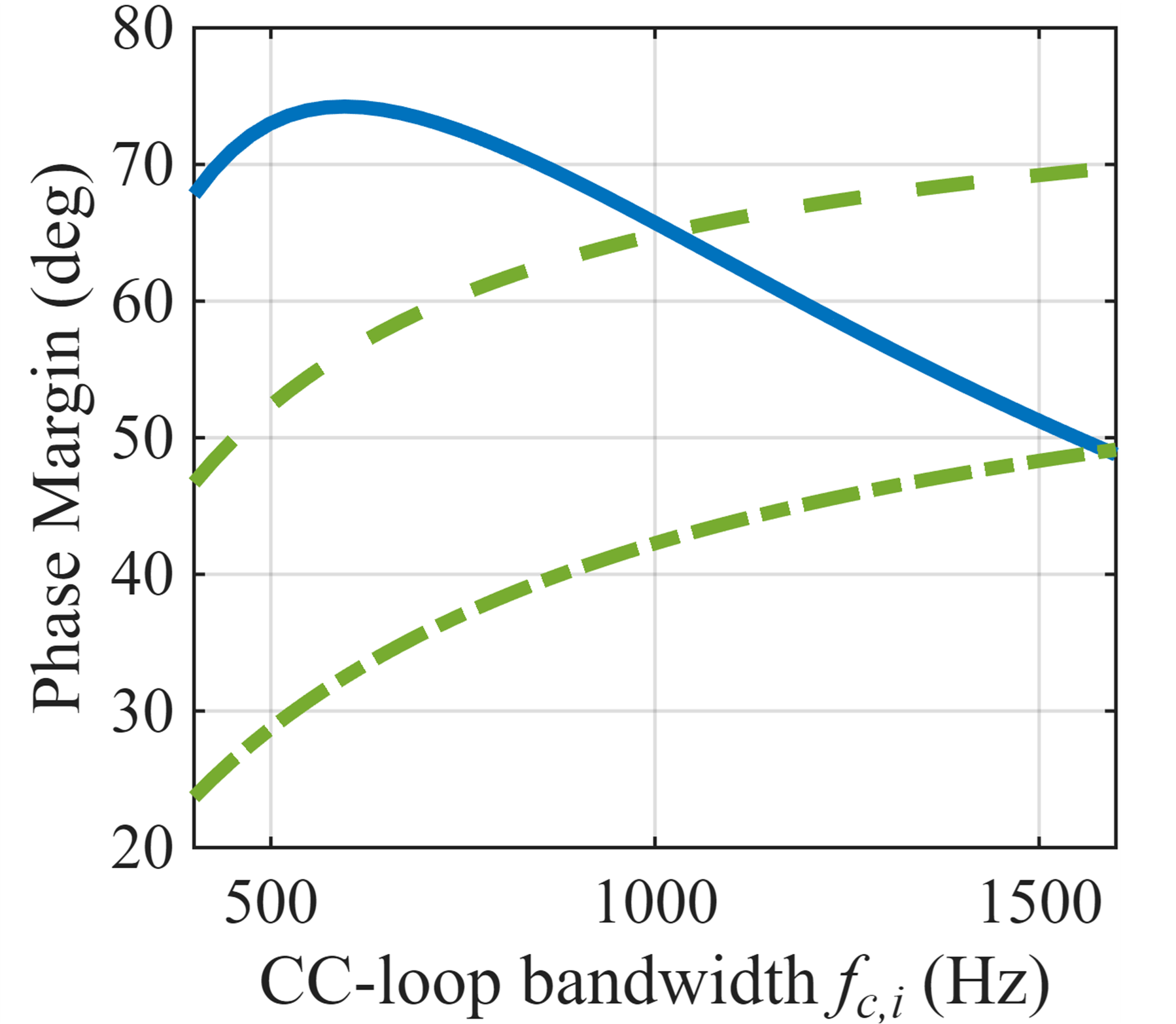}
        \label{fig:fig5b}}
    \caption{Full-model variation of VC-loop bandwidth and phase margin
    versus CC-loop bandwidth $f_{c,i}$. The baseline VA-CC GFMI is compared
    with unity VFD with AD.}
    \label{fig:fig5}
    \vspace{-15pt}
\end{figure}

Fig. \ref{fig:fig5} further clarifies the reduced dependence of $f_{c,v}$ on
the CC-loop bandwidth. The baseline VA-CC GFMI shows a direct increase in
$f_{c,v}$ as $f_{c,i}$ increases, confirming the bandwidth coupling predicted
by \eqref{eq:wc_baseline}. In contrast, unity VFD with AD keeps $f_{c,v}$
primarily determined by $K_{ad}$, with the expected phase-margin reduction
as $K_{ad}$ is decreased. Thus, Fig. \ref{fig:fig5} explicitly confirms the
design role of $K_{ad}$ as the main VC-loop shaping parameter, while $K_p$
mainly affects the achievable robustness margin.

\subsection{Implementation Refinement and Compact Design Equations}

The analysis in Section II used constant $K_{ad}$ to expose the
support-restoration mechanism. However, Fig. \ref{fig:fig3} shows that
constant AD also affects the fundamental component, which can degrade
steady-state tracking. Therefore, the implemented AD path is shaped by a
notch filter centered at $\omega_o$,

\begin{equation}
G_{\mathrm{NF}}(s)
=
\frac{s^2+\omega_o^2}
{s^2+2\zeta\omega_o s+\omega_o^2},
\label{eq:gnf}
\end{equation}
where $\zeta$ is the notch damping ratio, set to 0.1 in this study.

\begin{figure}[!t]
    \centering
    \includegraphics[width=0.95\columnwidth]{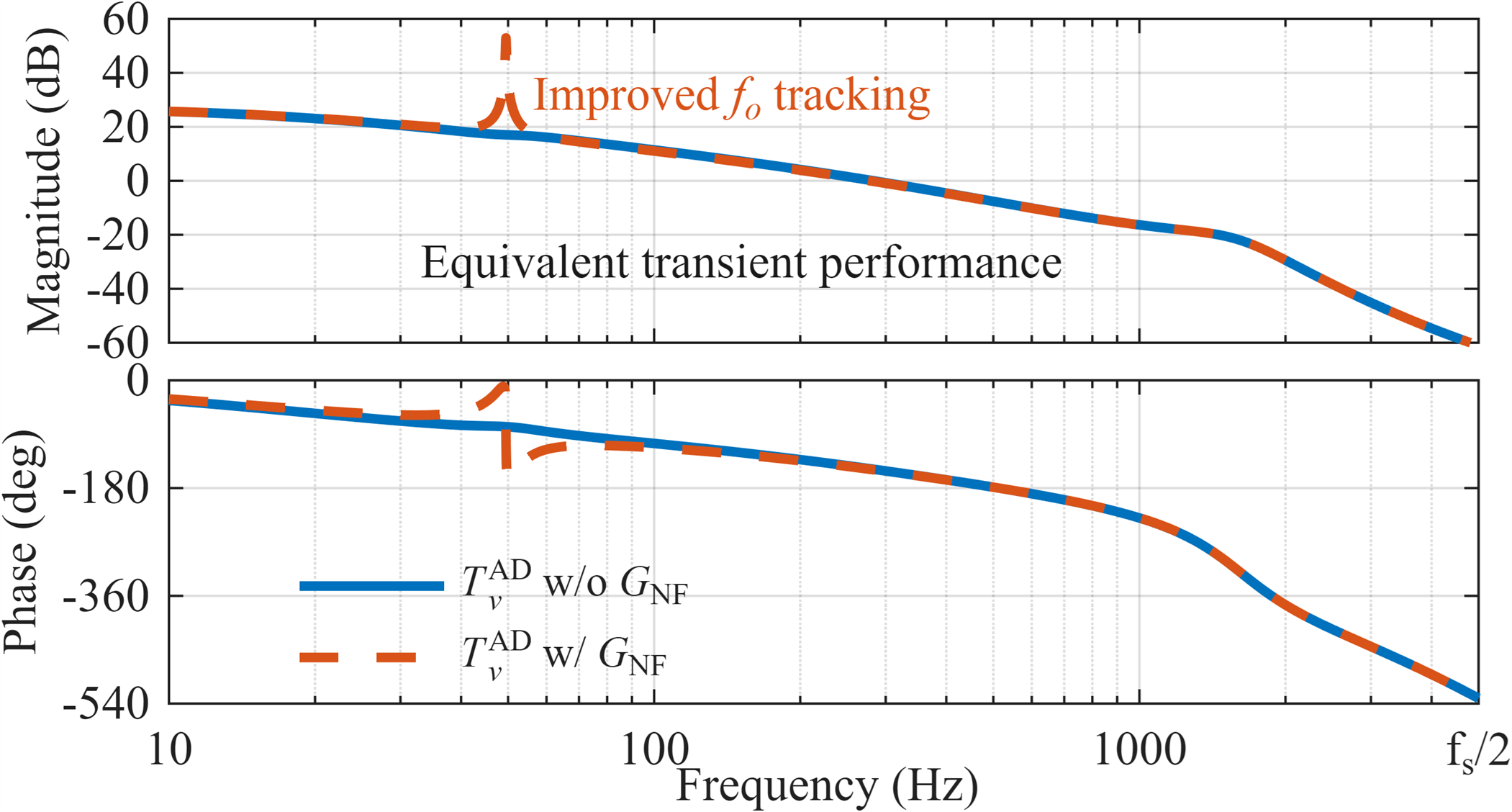}
    \caption{Open-loop Bode comparison of constant-AD and notch-filtered AD
    implementations.}
    \label{fig:fig6}
    \vspace{-15pt}
\end{figure}

Fig. \ref{fig:fig6} shows that the notch-filtered implementation preserves
the stabilizing effect of the AD path outside the notch band while reducing
its influence at $\omega_o$. Thus, the filtered AD path is treated as an
implementation refinement of the same support-restoration mechanism.

For design-oriented tuning, the task is to select $K_{ad}$ such that
sufficient phase margin is retained. To obtain closed-form expressions, the first-order delay model is combined
with the dominant filter-capacitor term; letting
$A_1 = C_fK_p + T_d(1-K_pK_{ad})$, the refined crossover estimate
(derived in the Appendix) is

\begin{equation}
\omega_{c,v}
\approx
\sqrt{
\frac{
-(K_pK_{ad})^2
+
\sqrt{(K_pK_{ad})^4+4A_1^2(K_p/L_v)^2}
}
{2A_1^2}
}.
\label{eq:wc_refined}
\end{equation}

The estimate in \eqref{eq:wc_refined} reduces to \eqref{eq:wc_ad} in the
low-frequency limit but captures the crossover saturation as the VC-loop
bandwidth approaches the trap-limited region. Solving \eqref{eq:wc_refined}
for $K_{ad}$ gives the inverse design law. Letting
$\beta=C_fK_p+T_d$,

\begin{equation}
K_{ad}
\approx
\frac{
\beta T_d\omega_{c,v}^3
+
\sqrt{
(K_p/L_v)^2(1+T_d^2\omega_{c,v}^2)
-\beta^2\omega_{c,v}^4
}
}
{
K_p\omega_{c,v}(1+T_d^2\omega_{c,v}^2)
}.
\label{eq:kad_inverse}
\end{equation}
A real solution requires the radicand in \eqref{eq:kad_inverse} to be
nonnegative, yielding the bandwidth-feasibility condition

\begin{equation}
K_p^2(1+T_d^2\omega_{c,v}^2)
-
L_v^2\beta^2\omega_{c,v}^4
\ge0 .
\label{eq:feasibility}
\end{equation}
Finally, with $A_1=C_fK_p+T_d(1-K_pK_{ad})$ as above, the compact
phase-margin estimate evaluated at the selected crossover frequency is

\begin{equation}
\mathrm{PM}
=
90^\circ
-
\left[
\omega_{c,v}T_d
+
\tan^{-1}
\left(
\frac{\omega_{c,v}A_1}{K_pK_{ad}}
\right)
\right]
\frac{180^\circ}{\pi}.
\label{eq:pm_compact}
\end{equation}
The derivations of the support condition and compact design relations are
provided in the Appendix.

\begin{figure}[!t]
    \centering
    \includegraphics[width=0.95\columnwidth]{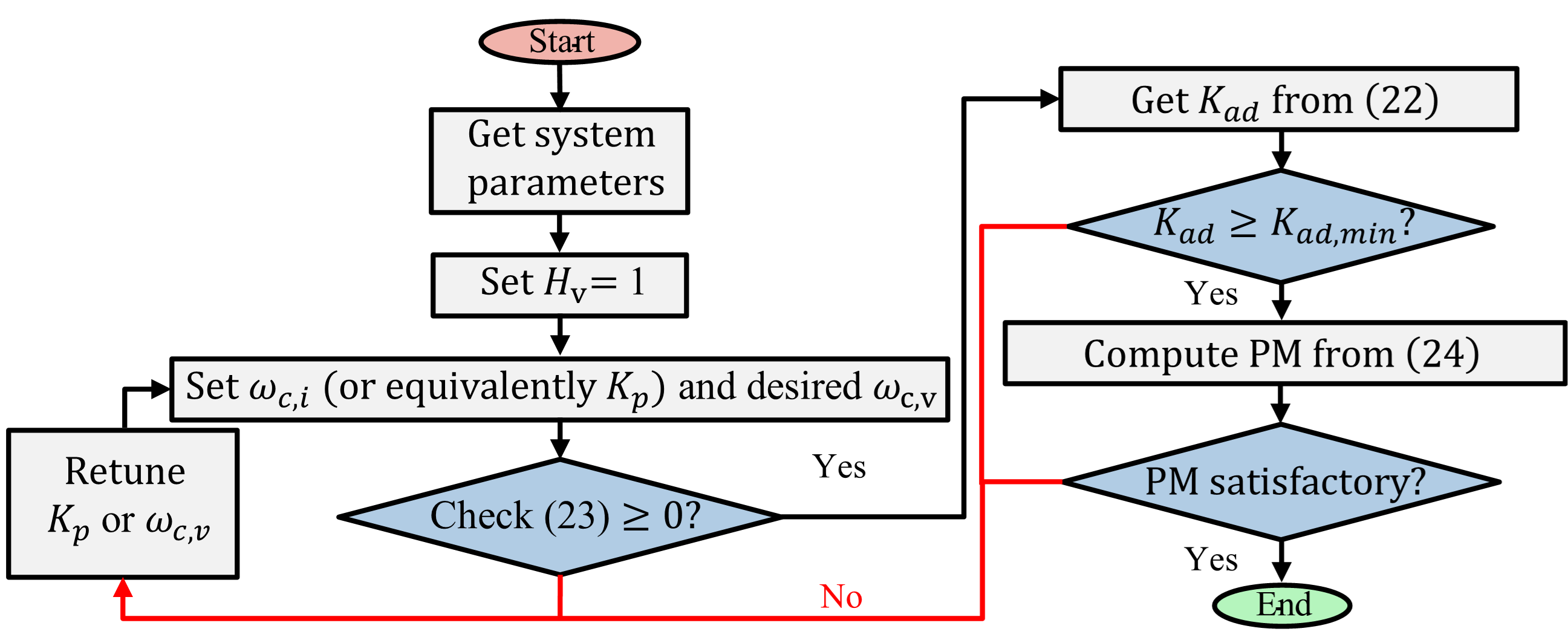}
    \caption{Design flowchart for selecting the AD gain from the target
    VC-loop bandwidth.}
    \label{fig:fig7}
    \vspace{-15pt}
\end{figure}

Taken together, \eqref{eq:kad_min} and
\eqref{eq:wc_refined}--\eqref{eq:pm_compact} provide the compact design tools
used in Fig. \ref{fig:fig7}: \eqref{eq:kad_min} sets the minimum support
requirement, \eqref{eq:kad_inverse} gives the initial $K_{ad}$ estimate for a
target bandwidth, \eqref{eq:feasibility} checks bandwidth feasibility,
\eqref{eq:wc_refined} predicts the achieved crossover, and
\eqref{eq:pm_compact} evaluates the resulting phase margin. The resulting
gain-selection procedure is summarized in Fig. \ref{fig:fig7}.

\subsection{Compact-Estimate Validation and Operating-Point Selection}

\begin{figure}[!t]
    \centering
    \subfloat[]{%
        \includegraphics[width=0.48\columnwidth]{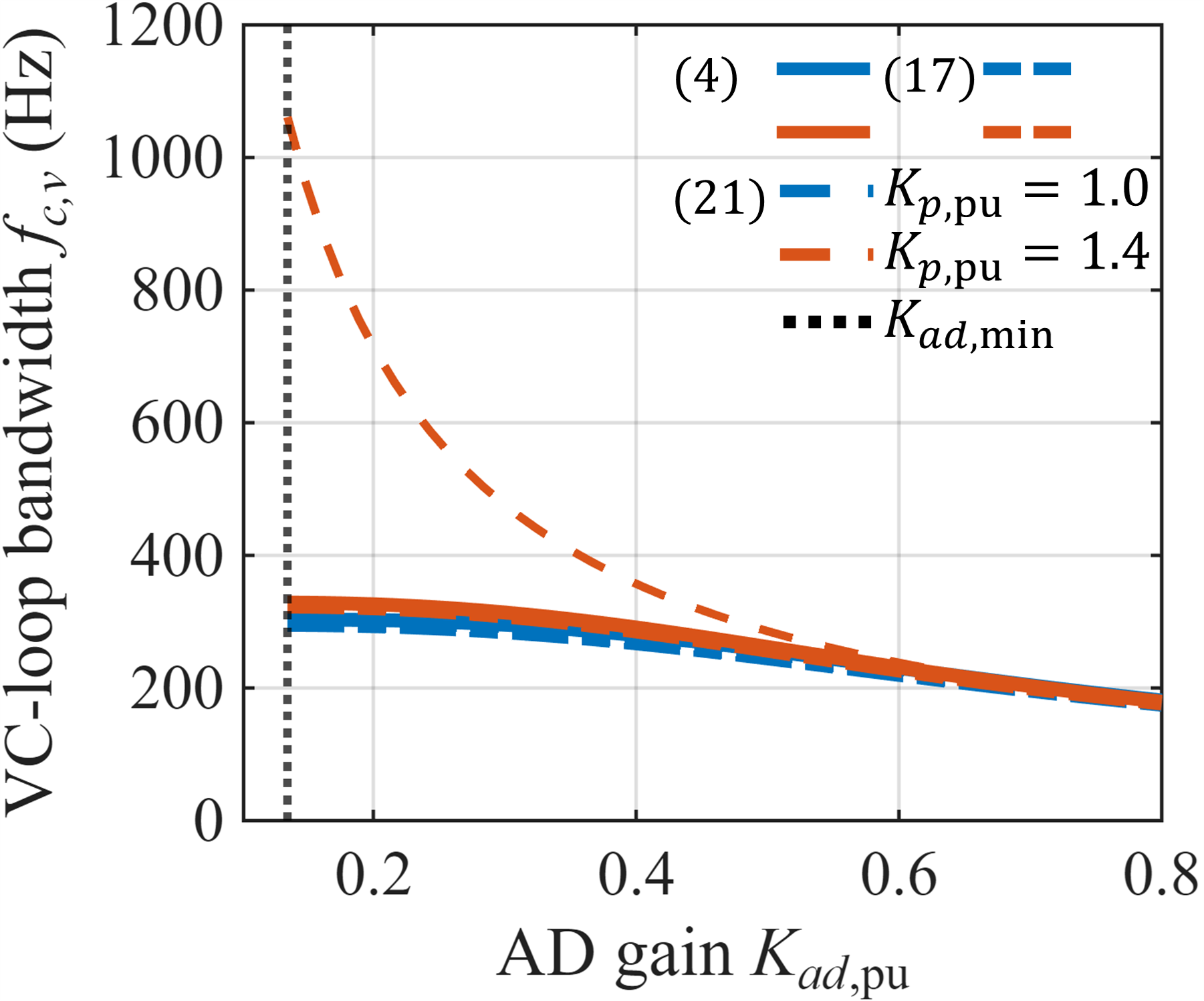}
        \label{fig:fig8a}}
    \hfill
    \subfloat[]{%
        \includegraphics[width=0.48\columnwidth]{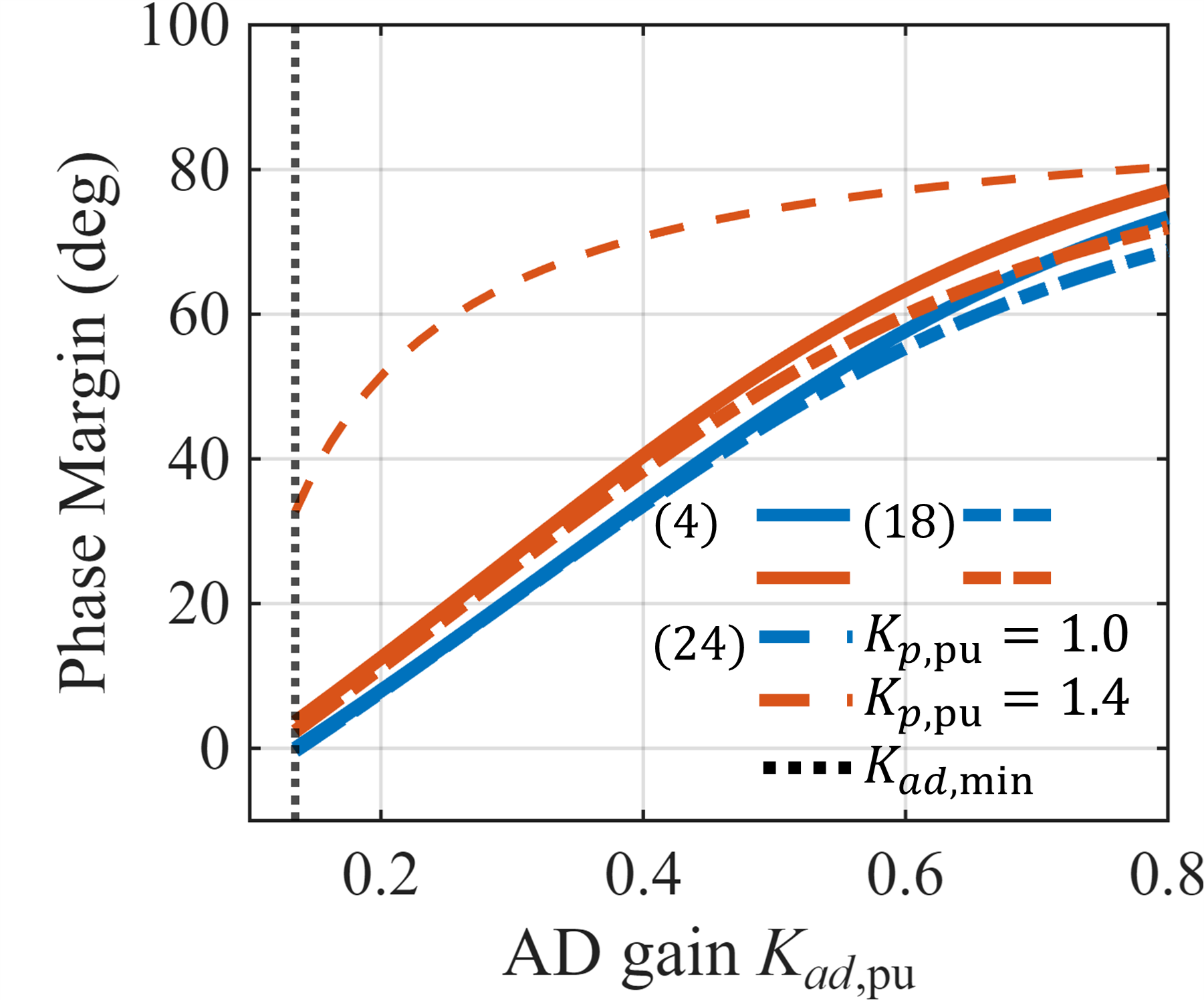}
        \label{fig:fig8b}}
    \caption{Compact-estimate validation against the corresponding full
    model: (a) VC-loop bandwidth and (b) phase margin.}
    \label{fig:fig8}
    \vspace{-15pt}
\end{figure}

Fig. \ref{fig:fig8} validates the compact estimates against the full-model
results. In Fig. \ref{fig:fig8a}, the leading-order relation in
\eqref{eq:wc_ad} captures the inverse $K_{ad}$--bandwidth trend only in the
low-bandwidth region, whereas the refined estimate in
\eqref{eq:wc_refined} follows the full-model trend and captures the
saturation near the trap region. Fig. \ref{fig:fig8b} similarly shows that
the leading-order PM in \eqref{eq:pm_ad} becomes optimistic near the
low-$K_{ad}$ boundary, while \eqref{eq:pm_compact} captures the phase-margin
collapse. Thus, the leading-order relations are useful for structural
interpretation, while the compact relations are used for operating-point
selection.

\begin{table}[!t]
\caption{Representative Operating Points for Validation}
\label{tab:operating_points}
\centering
\renewcommand{\arraystretch}{1.15}
\begin{tabular}{c c c c c c}
\toprule
\toprule
Case & $K_{ad,\mathrm{pu}}$ & $H_v$ & $f_{c,v}$ & PM ($^\circ$) & Status \\
\midrule
Baseline & 0 & 0 & 143 Hz & 82.3 & $\triangle$ \\
Nominal & 0.38 & 1 & 272 Hz & 31.5 & $\circ$ \\
Aggressive & 0.20 & 1 & 293 Hz & 8.4 & $\triangle$ \\
Insufficient & 0.08 & 1 & 294 Hz & $-6.4$ & $\times$ \\
\bottomrule
\bottomrule
\end{tabular}

\vspace{0.5ex}
\footnotesize
Note: Baseline PM from \eqref{eq:pm_baseline}; all other PM values from
\eqref{eq:pm_compact}. $\circ$ denotes the nominal selected point,
$\triangle$ denotes a stable but non-nominal reference point, and $\times$
denotes an insufficient-support point with negative stability margin.
\vspace{-10pt}
\end{table}

Based on the procedure in Fig. \ref{fig:fig7} and the compact estimates
validated in Fig. \ref{fig:fig8}, Table \ref{tab:operating_points} selects
four representative cases. The baseline reference is calculated from
\eqref{eq:wc_baseline} and \eqref{eq:pm_baseline}, which is appropriate
because its crossover lies in the low-bandwidth region where the
leading-order relation remains accurate. The nominal and aggressive
unity-VFD-with-AD cases are selected using \eqref{eq:kad_min} and
\eqref{eq:wc_refined}--\eqref{eq:pm_compact}, whereas the
insufficient-support case is intentionally chosen below the support boundary
in \eqref{eq:kad_min} to verify the predicted loss of stability. All selected
cases use $K_{p,\mathrm{pu}}=1.0$.

\begin{figure}[!t]
    \centering
    \includegraphics[width=0.95\columnwidth]{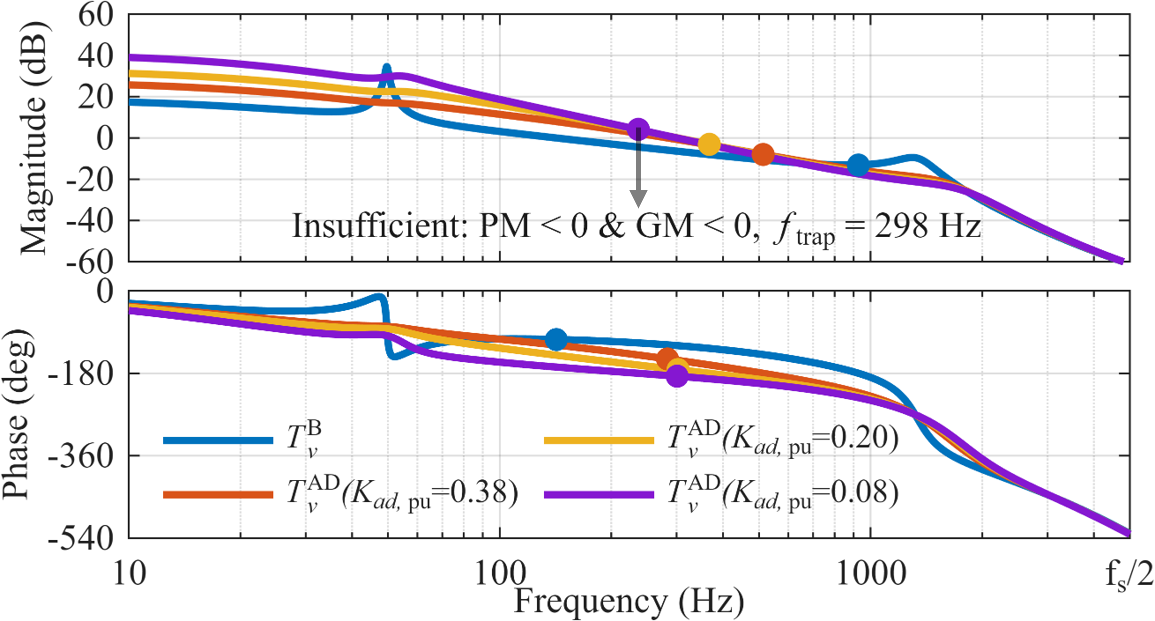}
    \caption{Open-loop Bode comparison of the selected operating points.}
    \label{fig:fig9}
    \vspace{-8pt}
\end{figure}

Fig. \ref{fig:fig9} verifies the selected cases using the full-model
open-loop Bode responses. The baseline case remains stable but
bandwidth-limited, the nominal case achieves the intended bandwidth increase
with usable phase margin, and the aggressive case approaches the
trap-limited region with reduced margin. The insufficient-support case
crosses the stability boundary near the trap region, consistent with the
negative compact phase margin in Table \ref{tab:operating_points}. Retained
LC/current-loop dynamics and exact delay slightly shift the full-model
crossovers from the compact estimates while preserving the predicted
stability ordering.

\section{Experimental Validation}

\begin{figure}[!t]
    \centering
    \includegraphics[width=0.95\columnwidth]{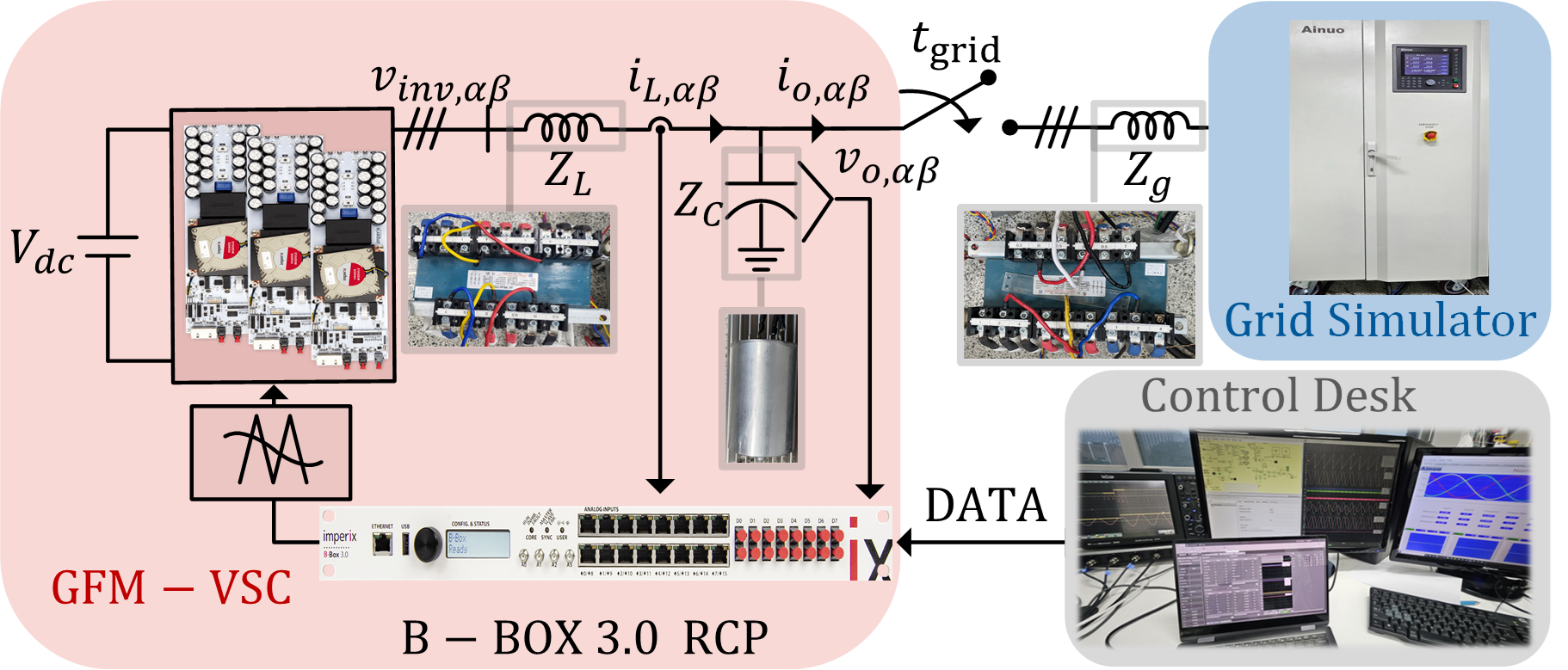}
    \caption{Experimental platform used for standalone and grid-connected
    validation.}
    \label{fig:fig10}
    \vspace{-15pt}
\end{figure}

Fig. \ref{fig:fig10} shows the experimental platform used for validation.
The setup consists of Imperix PEB 8038 power modules connected through the
LC filter, a B-Box RCP 3 controller for implementation of the considered
control architecture, a bidirectional dc power supply (IT6018T), a grid
simulator (ANBGS), and a host control computer. The same platform is used for
the standalone and grid-connected tests in Figs. \ref{fig:fig11}--\ref{fig:fig14}.

This section validates the proposed structural interpretation and design
framework experimentally. The unloaded standalone tests first verify the VFD
trap, the selected operating-point logic, and the $K_{ad}$-based VC-loop
tuning effect without external grid interaction. The grid-connected tests
then evaluate the final notch-filtered AD implementation under different
grid strengths and compare it with the baseline and constant-AD cases.

\subsection{Standalone Experimental Validation}

\begin{figure}[!t]
    \centering
    \subfloat[]{%
        \includegraphics[width=0.95\columnwidth]{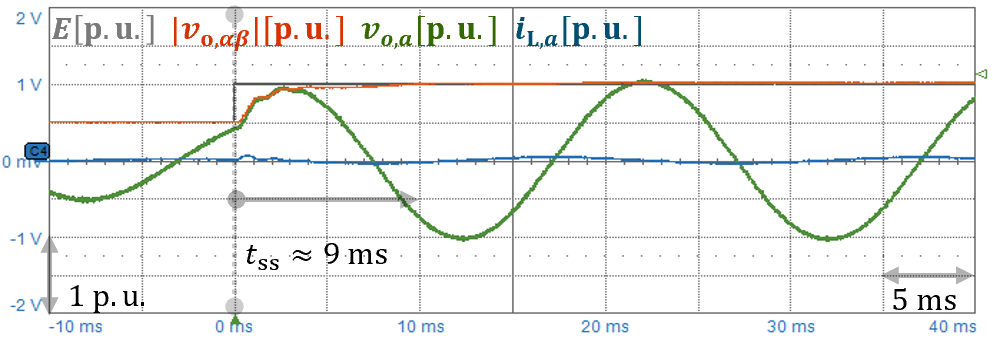}
        \label{fig:fig11a}}\\
    \subfloat[]{%
        \includegraphics[width=0.95\columnwidth]{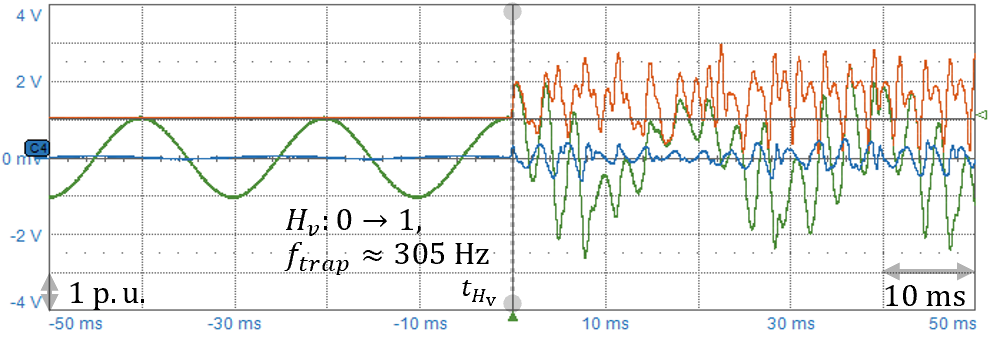}
        \label{fig:fig11b}}\\
    \subfloat[]{%
        \includegraphics[width=0.95\columnwidth]{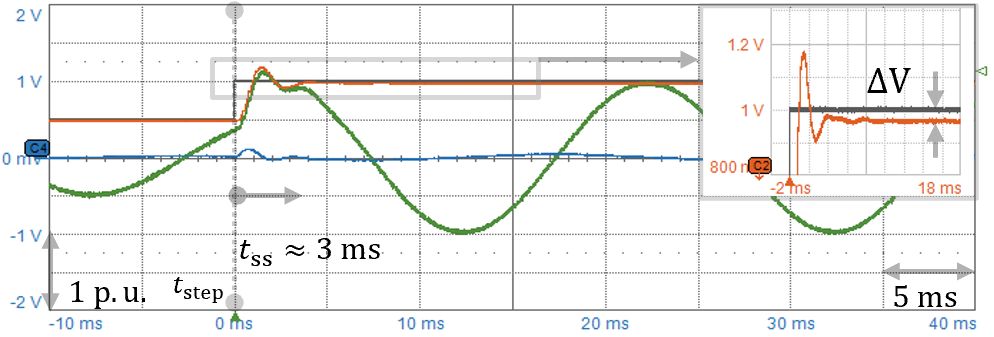}
        \label{fig:fig11c}}\\
    \subfloat[]{%
        \includegraphics[width=0.95\columnwidth]{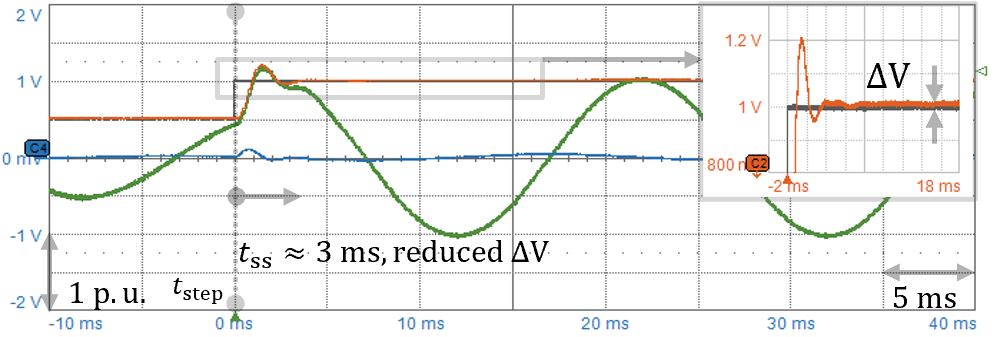}
        \label{fig:fig11d}}\\
    \subfloat[]{%
        \includegraphics[width=0.95\columnwidth]{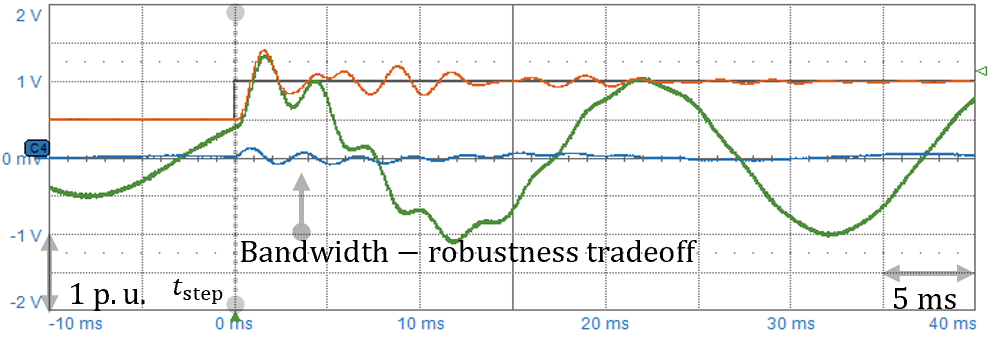}
        \label{fig:fig11e}}\\
    \subfloat[]{%
        \includegraphics[width=0.95\columnwidth]{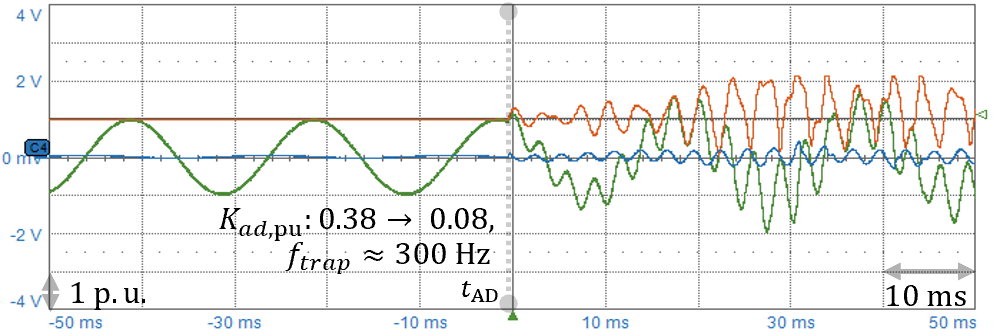}
        \label{fig:fig11f}}
    \caption{Standalone experimental responses under a 0.5--1-p.u.
    voltage-reference step and parameter-switching tests: (a) baseline
    VA-CC GFMI, (b) unity-VFD trap case, (c) nominal unity VFD with AD,
    (d) nominal unity VFD with notch-filtered AD, (e) aggressive case, and
    (f) insufficient-support case.}
    \label{fig:fig11}
    \vspace{-15pt}
\end{figure}

In Fig. \ref{fig:fig11}, $E$ and
$|v_{o,\alpha\beta}|=\sqrt{v_{o,\alpha}^{2}+v_{o,\beta}^{2}}$ represent the
reference voltage magnitude and output-voltage vector magnitude,
respectively, while $v_{o,a}$ and $i_{L,a}$ denote the $a$-phase output
voltage and $a$-phase filter-inductor current, respectively. The markers
$t_{\mathrm{step}}$, $t_{H_v}$, and $t_{\mathrm{AD}}$ denote the
voltage-reference step, VFD-gain switching instant, and AD-gain switching
instant, respectively. Here, $t_{\mathrm{ss}}$ denotes the 2\% settling time,
and $\Delta V$ denotes the steady-state magnitude difference between $E$ and
$|v_{o,\alpha\beta}|$.

Under a common voltage-reference step from 0.5 to 1 p.u.,
Fig. \ref{fig:fig11a} shows that the baseline VA-CC GFMI remains stable but
has a settling time of $t_{\mathrm{ss}}\approx9~\mathrm{ms}$. In contrast,
Fig. \ref{fig:fig11b} shows that switching $H_v$ from 0 to 1 at $t_{H_v}$
excites the unity-VFD trap mode at $f_{\mathrm{trap}}\approx305~\mathrm{Hz}$,
consistent with the unity-VFD mechanism in \eqref{eq:Tv_trap}--\eqref{eq:pm_trap}
and Fig. \ref{fig:fig3}. The deviation from the leading-order estimate in
\eqref{eq:wtrap} is expected because retained filter dynamics and delay shift
the actual trap region downward.

With the proposed AD path, Fig. \ref{fig:fig11c} shows that the nominal case
recovers stable operation and reduces the settling time to
$t_{\mathrm{ss}}\approx3~\mathrm{ms}$. Subsequently, Fig. \ref{fig:fig11d}
confirms that the notch-filtered implementation preserves the fast transient
response while reducing $\Delta V$. Finally, the boundary cases in
Figs. \ref{fig:fig11e} and \ref{fig:fig11f} clarify the two design limits:
aggressive tuning near the high-bandwidth edge reduces robustness, whereas
switching $K_{ad,\mathrm{pu}}$ from 0.38 to 0.08 at $t_{\mathrm{AD}}$ leads
to insufficient support and oscillation near 300 Hz.

\begin{figure}[!t]
    \centering
    \subfloat[]{%
        \includegraphics[width=0.48\columnwidth]{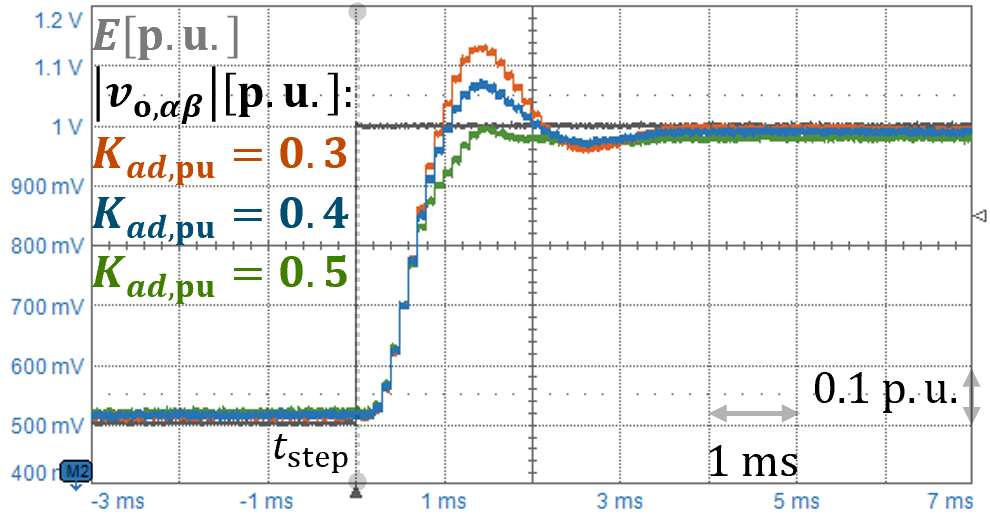}
        \label{fig:fig12a}}
    \hfill
    \subfloat[]{%
        \includegraphics[width=0.48\columnwidth]{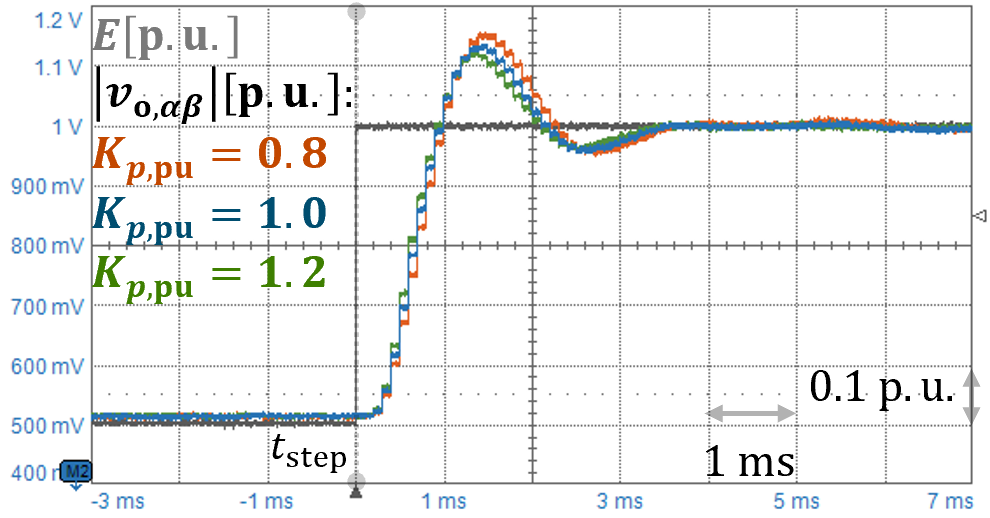}
        \label{fig:fig12b}}
    \caption{Experimental validation of VC-loop tuning under unity VFD with
    notch-filtered AD: (a) $K_{ad}$ sweep at $K_{p,\mathrm{pu}}=1.0$ and
    (b) $K_p$ sweep at $K_{ad,\mathrm{pu}}=0.3$.}
    \label{fig:fig12}
    \vspace{-5pt}
\end{figure}

Fig. \ref{fig:fig12} further verifies the VC-loop tuning behavior of the
proposed AD path using the voltage-magnitude response to the same reference
step at $t_{\mathrm{step}}$. In Fig. \ref{fig:fig12a}, increasing
$K_{ad,\mathrm{pu}}$ from 0.3 to 0.5 at fixed $K_{p,\mathrm{pu}}=1$ reduces
the refined VC-loop crossover from approximately 290 to 250 Hz while
improving transient damping. In contrast, Fig. \ref{fig:fig12b} shows that
increasing $K_{p,\mathrm{pu}}$ from 0.8 to 1.2 at fixed
$K_{ad,\mathrm{pu}}=0.3$ produces only a moderate transient-shaping effect.
Therefore, $K_{ad}$ acts as the dominant VC-loop tuning variable, whereas
$K_p$ provides only a secondary correction, consistent with the weak
$f_{c,i}$-dependence in Fig. \ref{fig:fig5}.

\subsection{Grid-Connected Experimental Validation}

\begin{figure}[!t]
    \centering
    \subfloat[]{%
        \includegraphics[width=0.95\columnwidth]{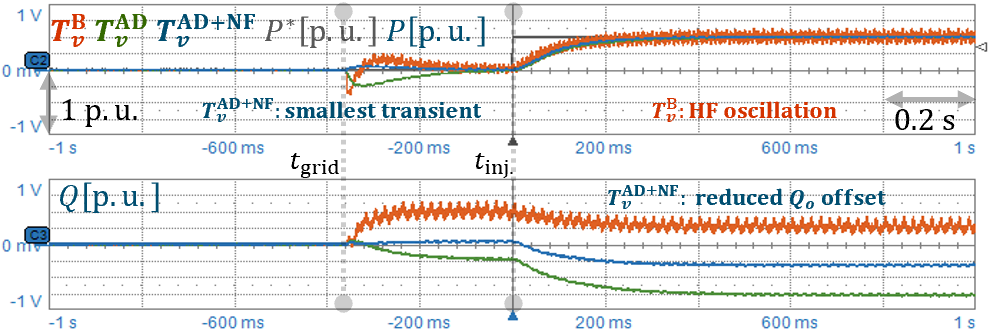}
        \label{fig:fig13a}}\\
    \subfloat[]{%
        \includegraphics[width=0.95\columnwidth]{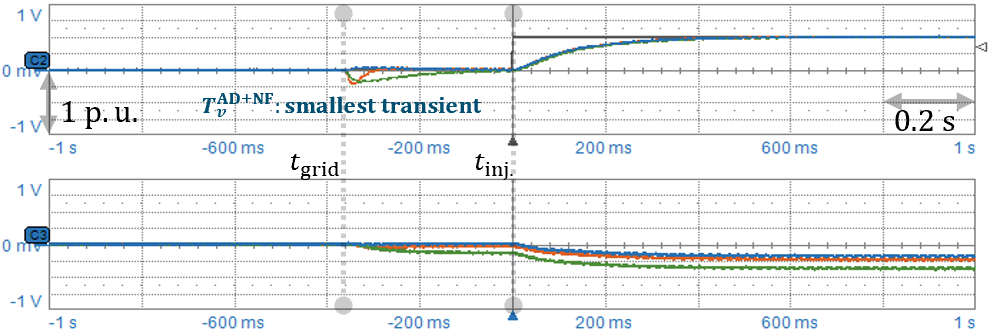}
        \label{fig:fig13b}}
    \caption{Grid-connected comparison at 0.5-p.u. active-power injection
    under different grid strengths: (a) SCR = 20 and (b) SCR = 3. The
    baseline VA-CC GFMI, unity VFD with constant AD, and unity VFD with
    notch-filtered AD are compared.}
    \label{fig:fig13}
    \vspace{-15pt}
\end{figure}

\begin{figure}[!t]
    \centering
    \subfloat[]{%
        \includegraphics[width=0.95\columnwidth]{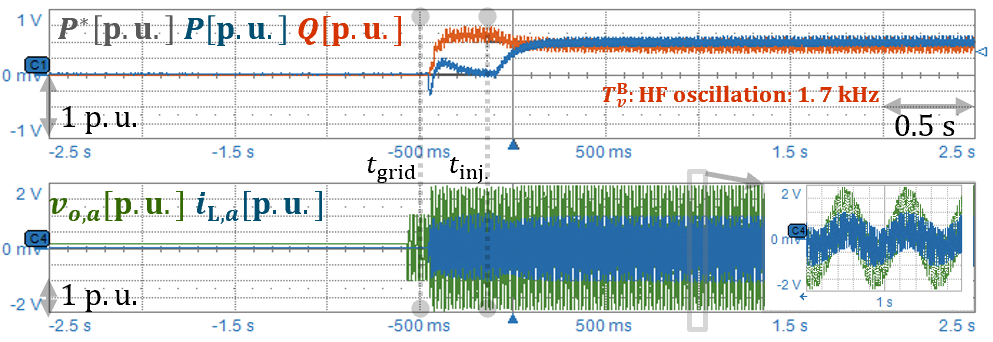}
        \label{fig:fig14a}}\\
    \subfloat[]{%
        \includegraphics[width=0.95\columnwidth]{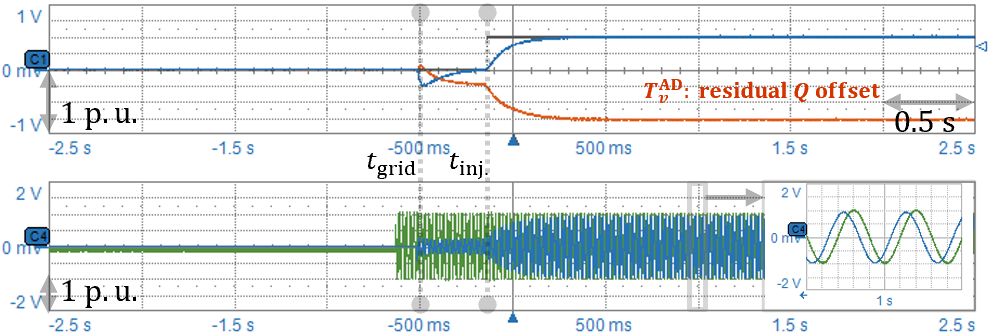}
        \label{fig:fig14b}}\\
    \subfloat[]{%
        \includegraphics[width=0.95\columnwidth]{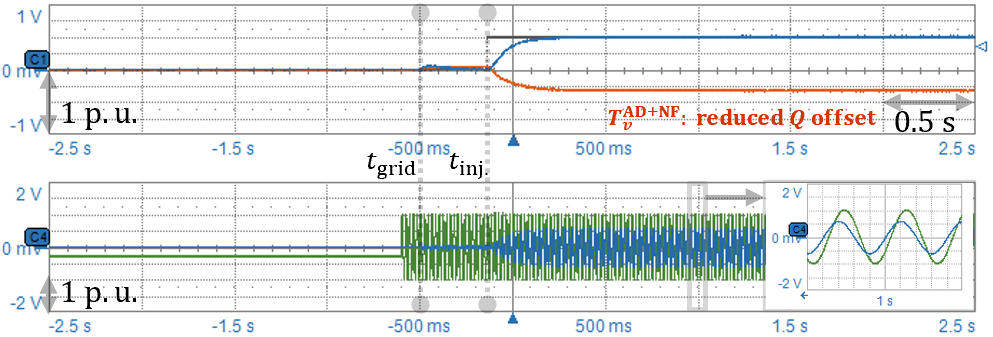}
        \label{fig:fig14c}}
    \caption{Grid-connected controller comparison at SCR = 20: (a) baseline
    VA-CC GFMI, (b) nominal unity VFD with constant AD, and (c) nominal unity
    VFD with notch-filtered AD.}
    \label{fig:fig14}
\vspace{-15pt}
\end{figure}

Grid-connected operation is evaluated at 0.5-p.u. active power. In
Figs. \ref{fig:fig13} and \ref{fig:fig14}, $t_{\mathrm{grid}}$ and
$t_{\mathrm{inj}}$ denote the grid-connection instant and active-power
injection instant, respectively. Fig. \ref{fig:fig13} compares the baseline
VA-CC GFMI, unity VFD with AD, and unity VFD with notch-filtered AD under
SCR = 20 and SCR = 3. Under SCR = 20, the baseline VA-CC case exhibits
high-frequency oscillation and a large $Q$ offset after $t_{\mathrm{inj}}$,
whereas the notch-filtered AD case reduces both the transient deviation and
the $Q$ offset. Under SCR = 3, the notch-filtered AD case also maintains
stable power injection, indicating that the proposed AD path remains
effective under the tested weak-grid condition.

Fig. \ref{fig:fig14} further compares the detailed grid-connected waveforms
at SCR = 20. The baseline case exhibits high-frequency oscillation after
$t_{\mathrm{inj}}$, whereas both AD-based cases recover stable operation.
Compared with constant AD, the notch-filtered implementation reduces the
$Q$ offset while retaining stable active-power tracking and stable
voltage-current waveforms. These results indicate that the notch-filtered
AD path preserves the transient benefit of the proposed AD mechanism while
improving the steady-state grid-connected behavior.

\section{Conclusion}

This paper presented a structural analysis and internal-stability enhancement
method for VA-based GFMIs under unity VFD. The analysis showed that
unity VFD can relax the direct VC-loop bandwidth dependence on the CC loop,
but it also removes the low-frequency restoring term associated with the
filter capacitor. In the presence of digital control delay, this drives the
VC loop toward a delay-sensitive double-integrator structure, identified as
the VFD trap.

To recover internal stability without attenuating VFD, a proportional AD path
was proposed as negative capacitor-voltage feedback in the current-reference
path. The proposed method restores the low-frequency support removed by unity
VFD and provides $K_{ad}$ as a practical VC-loop tuning parameter. A minimum
support condition, delay-aware phase-margin estimate, and refined
forward/inverse design equations were derived for operating-point selection.
Analytical and experimental results verified stable unity-VFD operation,
improved VC-loop bandwidth, and predictable tuning behavior across nominal,
aggressive, and insufficient-support cases.

Future work could extend the proposed design to loaded and grid-connected
impedance models so that the interaction among VA parameters, AD tuning, and
grid strength can be incorporated into the closed-form design process.

\appendices
\renewcommand{\theequation}{A\arabic{equation}}
\setcounter{equation}{0}
\section{Derivation of the Support and Design Conditions}

The main text uses reduced closed-form relations to expose the
support-restoration mechanism and to select practical AD operating points.
This appendix summarizes the algebra leading to the minimum support
condition, the compact crossover estimate, the inverse $K_{ad}$ design law,
and the phase-margin expression. The derivation starts from the AD-modified
inner-loop denominator and applies the same first-order delay approximation
used in the main analysis.

For unity VFD with AD, the AD-modified inner-loop denominator is

\begin{equation}
\begin{aligned}
\Delta_{ad}(s)
&=
s^2L_fC_f+sC_fG_d(s)G_i(s)  \\
&\quad
+1-G_d(s)+G_d(s)G_i(s)K_{ad}.
\end{aligned}
\label{eq:app_delta_ad_full}
\end{equation}
With $G_i(s)\approx K_p$ and $G_d(s)\approx1-sT_d$, its low-frequency form
becomes

\begin{equation}
\Delta_{ad}(s)\approx K_pK_{ad}+sA_1,
\label{eq:app_delta_ad_lf}
\end{equation}
where

\begin{equation}
A_1=C_fK_p+T_d(1-K_pK_{ad}).
\label{eq:app_A1}
\end{equation}

The cleared VC characteristic expression is

\begin{equation}
\Delta_v(s)=sL_v\Delta_{ad}(s)+G_d(s)K_p .
\label{eq:app_delta_v}
\end{equation}
Substituting \eqref{eq:app_delta_ad_lf} into \eqref{eq:app_delta_v} and
retaining the low-frequency terms gives

\begin{equation}
\Delta_v(s)
\approx
K_p+s(L_vK_pK_{ad}-K_pT_d)+s^2L_vA_1 .
\label{eq:app_delta_v_lf}
\end{equation}

Then, the support boundary is obtained by setting the first-order coefficient in
\eqref{eq:app_delta_v_lf} to zero:

\begin{equation}
L_vK_pK_{ad}-K_pT_d=0,
\label{eq:app_support_boundary}
\end{equation}
which gives

\begin{equation}
K_{ad,\mathrm{min}}=\frac{T_d}{L_v}.
\label{eq:app_kad_min}
\end{equation}

The same reduced denominator gives the compact VC-loop transfer function

\begin{equation}
T_v^{\mathrm{AD}}(s)
\approx
\frac{G_d(s)K_p}
{sL_v(K_pK_{ad}+sA_1)} .
\label{eq:app_Tv_ad}
\end{equation}
At $s=j\omega_{c,v}$, the magnitude condition
$|T_v^{\mathrm{AD}}(j\omega_{c,v})|=1$ yields

\begin{equation}
A_1^2\omega_{c,v}^4
+
(K_pK_{ad})^2\omega_{c,v}^2
-
\left(\frac{K_p}{L_v}\right)^2
=0.
\label{eq:app_mag_condition}
\end{equation}

Solving \eqref{eq:app_mag_condition} for the positive $\omega_{c,v}$ gives
\eqref{eq:wc_refined}. For inverse design, using

\begin{equation}
\beta=C_fK_p+T_d,
\qquad
A_1=\beta-T_dK_pK_{ad},
\label{eq:app_beta}
\end{equation}
and solving \eqref{eq:app_mag_condition} for $K_{ad}$ gives
\eqref{eq:kad_inverse}. The real-solution requirement gives the feasibility
condition in \eqref{eq:feasibility}.

Finally, the phase of \eqref{eq:app_Tv_ad} is approximated as

\begin{equation}
\angle T_v^{\mathrm{AD}}(j\omega_{c,v})
\approx
-\omega_{c,v}T_d
-90^\circ
-\tan^{-1}
\left(
\frac{\omega_{c,v}A_1}{K_pK_{ad}}
\right),
\label{eq:app_phase}
\end{equation}
and using
$\mathrm{PM}=180^\circ+\angle T_v^{\mathrm{AD}}(j\omega_{c,v})$
gives \eqref{eq:pm_compact}.

The virtual resistance $R_v$ is neglected in the closed-form structural
derivations because the relevant VC-loop crossover and trap frequencies 
are well above the virtual-admittance corner frequency $R_v/L_v$. 
Retaining $R_v$ adds a small damping contribution and
slightly lowers the support floor, but does not change the unity-VFD
support-loss mechanism.

\balance
\bibliographystyle{IEEEtran}
\bibliography{vfd_trap_references}


\end{document}